# Initial PET performance evaluation of a preclinical insert for PET/MRI with digital SiPM technology

David Schug[1], Christoph Lerche[2,3], Bjoern Weissler[1,4], Pierre Gebhardt[5], Benjamin Goldschmidt[1], Jakob Wehner[1], Peter Michael Dueppenbecker[4,5], Andre Salomon[2], Patrick Hallen[1], Fabian Kiessling[6] and Volkmar Schulz[1,4]

[1] Physics of Molecular Imaging Systems, Experimental Molecular Imaging, RWTH Aachen University, 52062 Aachen, Germany
[2] Oncology Solutions, Philips Research, 5656 AE Eindhoven, Netherlands
[3] Institute of Neuroscience and Medicine (INM-4), Forschungszentrum Jülich GmbH, 52428 Jülich, Germany
[4] Clinical Application Research, Philips Research, 52068 Aachen, Germany
[5] Imaging Sciences and Biomedical Engineering, King's College London, London WC2R 2LS, UK
[6] Experimental Molecular Imaging, RWTH Aachen University, 52062 Aachen, Germany

E-mail: schug@pmi.rwth-aachen.de and schulz@pmi.rwth-aachen.de

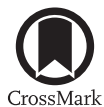

## Abstract

Hyperion-II$^D$ is a positron emission tomography (PET) insert which allows simultaneous operation in a clinical magnetic resonance imaging (MRI) scanner. To read out the scintillation light of the employed lutetium yttrium orthosilicate crystal arrays with a pitch of 1 mm and 12 mm in height, digital silicon photomultipliers (*DPC 3200-22*, Philips Digital Photon Counting) (DPC) are used. The basic PET performance in terms of energy resolution, coincidence resolution time (CRT) and sensitivity as a function of the operating parameters, such as the operating temperature, the applied overvoltage, activity and configuration parameters of the DPCs, has been evaluated at system level. The measured energy resolution did not show a large dependency on the selected parameters and is in the range of 12.4%–12.9% for low activity, degrading to ~13.6% at an activity of ~100 MBq. The CRT strongly depends on the selected trigger scheme (trig) of the DPCs, and we measured approximately 260 ps, 440 ps, 550 ps and 1300 ps for trig 1–4, respectively. The trues sensitivity for a NEMA NU 4 mouse-sized scatter phantom with a 70 mm long tube of activity was dependent on the operating parameters and was determined to be 0.4%–1.4% at low activity. The random fraction stayed below 5% at activity up to 100 MBq and the scatter fraction was evaluated as ∼6% for an energy window of 411 keV–561 keV and ∼16% for 250 keV–625 keV. Furthermore, we performed imaging experiments using a mouse-sized hot-rod phantom and a large rabbit-sized phantom. In 2D slices of the reconstructed mouse-sized hot-rod phantom ($\varnothing = 28$ mm), the rods were distinguishable from each other down to a rod size of 0.8 mm. There was no benefit from the better CRT of trig 1 over trig 3, where in the larger rabbit-sized phantom ($\varnothing = 114$ mm) we were able to show a clear improvement in image quality using the time-of-flight information. The findings will allow system architects—aiming at a similar detector design using DPCs—to make predictions about the design requirements and the performance that can be expected.

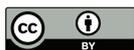

Online supplementary data available from stacks.iop.org/PMB/61/2851/mmedia

(Some figures may appear in colour only in the online journal)

## 1. Introduction

Positron emission tomography (PET) is a very sensitive functional imaging modality, e.g. for metabolic processes, but it provides almost no anatomical information. To allow anatomical co-registration, PET was successfully integrated with X-ray computed tomography (CT). However, PET/CT exposes the patients to an additional radiation dose, offers only limited soft tissue contrast and does not allow simultaneous imaging. The combination of PET with magnetic resonance imaging (MRI) promises to overcome these shortcomings (von Schulthess and Schlemmer 2009, Buchbender *et al* 2012a, 2012b, Drzezga *et al* 2012, Jadvar and Colletti 2014), but designing a PET/MRI system is challenging due to the high static magnetic field, the fast switching gradient fields and the radio frequency (RF) system needed for MR image acquisition (Vandenberghe and Marsden 2015). PET/MRI offers real simultaneous image acquisition, provided that the interference between both modalities is reduce to a tolerable level.

  Until recently, conventional PET detectors have been based on photomultiplier tubes which cannot be operated inside strong magnetic fields, and, therefore, silicon-based photo detectors are commonly used for PET/MRI applications. An overview of designs that combine PET with MRI can be found in Disselhorst *et al* (2014), Pichler *et al* (2008), Vandenberghe and Marsden (2015) and Zaidi and Del Guerra (2011).

  Analog silicon photomultipliers (SiPMs) are made up of an array of highly sensitive single photon avalanche diodes (SPAD), which give an analog avalanche signal for each breakdown. All the SPADs are coupled together and the sum of the current signal is proportional to the number of them that break down. The analog SiPM signal output is influenced by voltage and temperature changes and depends on photon detection efficiency (PDE) as well as the gain of individual SPADs.

Digital SiPMs (dSiPMs) are the latest evolutionary step in silicon-based photo detectors. They digitize and actively quench the breakdown of individual single SPADs, and the digitized breakdowns are summed up digitally and can be handled using digital signal processing techniques. The number of SPAD breakdowns, the dSiPM signal output, is mainly influenced by changes in the SPAD PDE and not by the SPAD gain—as long as the breakdown is detected (Frach *et al* 2009). Furthermore, dSiPMs are good candidates for building up robust PET systems that can be operated inside an MRI environment, as no analog signal transmission lines before digitization are needed. In 2009, Philips digital photon counting (PDPC) presented the first implementation of a dSiPM—the so-called digital photon counter (DPC) (Degenhardt *et al* 2009, Frach *et al* 2009).

So far, investigation of the DPC for PET applications has mainly been conducted using small demonstrators based on the technology evaluation kit (TEK) provided by PDPC. For example, the influence of DPC configuration parameters on the read-out probability (Tabacchini *et al* 2014) and on the number of counted photons (van Dam *et al* 2012) was investigated, and the timing performance of TEK-based gamma detectors was evaluated (van Dam *et al* 2013). Furthermore, scintillator arrangements ranging from pixelated (Schug *et al* 2012, 2013, Yeom *et al* 2013, Georgiou *et al* 2014, Marcinkowski *et al* 2014) to monolithic scintillators (Seifert *et al* 2013) were implemented on the TEK. However, only a few imaging-capable demonstrators using DPCs on the TEK have been presented (España *et al* 2014, Schneider *et al* 2015). Apart from a single demonstrator built up by PDPC (Degenhardt *et al* 2012), the results presented for a clinical PET/CT (Miller *et al* 2014) and a scanner using the same readout platform as used in this work employing a clinical scintillator configuration (Schug *et al* 2015c), DPCs have not yet been investigated at larger system levels. The presented system, developed by our group, was used to perform the first extensive performance study of a DPC-based preclinical high-resolution PET using a scintillator readout with light sharing.

We developed the first MRI-compatible PET insert, called Hyperion-II$^D$, on the basis of DPCs installed on an MRI-compatible readout infrastructure designed by our group (Weissler *et al* 2012, 2015). The predecessor, based on similar architecture employing analog SiPMs with ASICs, Hyperion-I, is discussed in Weissler *et al* (2014). For both versions of the platform, the complete digitization is performed inside the MRI and digital information is sent out via optical Ethernet links to a data acquisition server. The first results of the PET/MRI interference of the Hyperion-II$^D$ platform using DPCs were presented in Schug *et al* (2015a), Wehner *et al* (2014) and Weissler *et al* (2015), and a detailed interference study was presented in Wehner *et al* (2015). Two hardware versions of the Hyperion-II$^D$ platform equipped with a single ring of clinical scintillators have recently been compared and tested for MR compatibility (Schug *et al* 2015c).

Hyperion-II$^D$ employs sensor tiles which are specially optimized for MR compatibility, but similar in terms of the geometrical layout to the sensor tiles distributed by PDPC. The scanner houses 60 sensor tiles, while the TEK platform can be used to read out a maximum of four tiles with the Tile TEK and eight tiles with the Module TEK, and thus offers only a limited maximum sensitivity and bore size, and limits PET performance evaluations to small system designs. In this work, we investigated the influence of the DPC operating parameters on the PET performance of the insert equipped with a preclinical scintillator configuration outside an MRI system. The used data processing and calibration techniques were presented in Schug *et al* (2015b) and can be used on a wide range of system designs based on DPCs utilizing similar scintillator geometries. We evaluated the influence of operating parameters—such as temperature and the applied voltage—as well as DPC configuration parameters on the energy resolution, the timing performance, the sensitivity and the spatial image resolution of the insert. The findings shall allow system architects—aiming at a similar detector design using

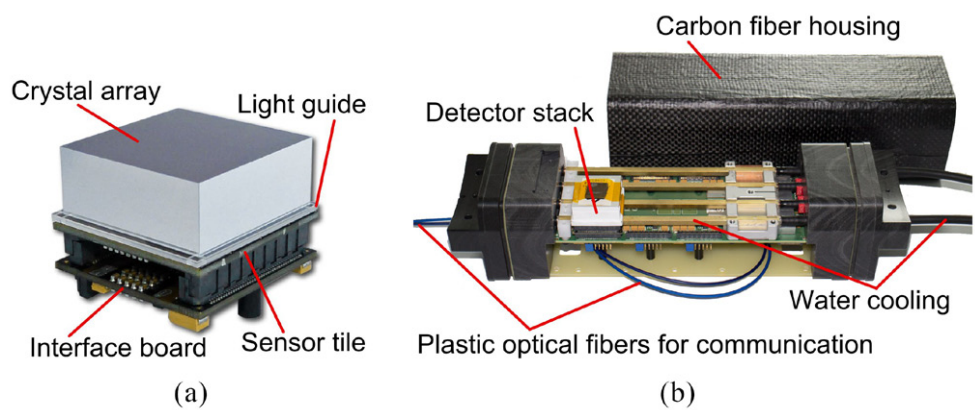

**Figure 1.** (a) The detector stack consists of an interface board, the sensor tile and a pixelated scintillating crystal array mounted on a light guide (© [2012] IEEE. Reprinted, with permission, from Weissler *et al* (2012)). (b) The singles detection module is able to house up to six stacks. Only one detector stack is mounted to reveal the cooling pipes (reprinted from Wehner *et al* (2014)).

DPCs—to make predictions about the design requirements and the performance that can be expected.

## 2. Materials

### 2.1. Hyperion-II$^D$ PET insert

The Hyperion-II$^D$ platform is described in detail in Weissler *et al* (2012) and Weissler *et al* (2015), and in Schug *et al* (2015b), we give a detailed overview of the detector components that are relevant for gamma detection.

The detector stack consists of a pixelated scintillating crystal array coupled via a light guide to a sensor tile, which is used to read out the scintillation light, and an FPGA-based control and readout board (interface (IF) board) (Dueppenbecker *et al* 2012b, 2016) (figure 1(a)). Optically isolated $30 \times 30$ cerium-doped lutetium yttrium orthosilicate (LYSO) crystals with a size of $0.933 \times 0.933 \times 12$ mm$^3$ and a pitch of 1 mm are mounted on a 2 mm thick glass light guide. The sensor tile is $32.6 \times 32.6$ mm$^2$ in size and is made up of 16 DPC 3200-22 sensors from PDPC with $2 \times 2$ pixels each (Degenhardt *et al* 2009, Frach *et al* 2009, Degenhardt *et al* 2010, Frach *et al* 2010), resulting in $8 \times 8$ readout channels per sensor tile. The 3200 SPADs per DPC pixel can be deactivated individually in order to reduce the overall dark count rate (DCR) of the DPC.

A trigger scheme (trig) defines how the trigger signal of four sub-regions of a pixel are logically connected to generate a trigger and can be set between 1–4 resulting in 1, $2.33 \pm 0.67$, $3.0 \pm 1.4$ and $8.33 \pm 3.80$ mean number of SPAD breakdowns per pixel in order to generate a trigger, respectively. Setting the trig to low values results in a better timing performance, but increases the likelihood of generating a trigger on dark counts, which may result in a higher dead time and lower sensitivity. In addition, the row-trigger-line refresh feature can help to reduce dark-noise-induced triggers and is used throughout this work (for details and explanation see Schug *et al* (2015b) and Marcinkowski *et al* (2013)).

When the trigger condition is met, a time stamp is generated and the DPC enters a validation phase of programmable length. The configurable validation scheme (val) is realized in a similar way to the trig, with a higher granularity and dynamic range, using eight bits to determine the threshold of SPAD breakdowns which has to be reached during the validation phase. The validation threshold can be set between 1 and $132 \pm 40$ SPAD breakdowns per pixel. Then, after validation, the integration phase is started, the DPC sums up the SPAD breakdowns and the hit data is transmitted. If a trigger is not validated, the integration phase is skipped, no hit data is transmitted, the DPC goes through a reset and becomes sensitive again. Details of the trig and val and the resulting thresholds can be found in Tabacchini *et al* (2014) and Philips Digital Photon Counting (2014), and the vals used in this work are listed in table 1.

**Table 1.** DPC-3200-22 validation schemes in the notation used throughout the text, hexadecimal notation, the notation used by PDPC and the resulting validation threshold per pixel (taken from Thon (2012), partly published in Philips Digital Photon Counting (2014) and Tabacchini *et al* (2014)).

| Val(text) | Val(hex) | Val(PDPC) | Avg. SPADs | Min. SPADs |
|---|---|---|---|---|
| 17ph | 0x55:OR | 4-OR | $16.9 \pm 6.2$ | 4 |
| 28ph | 0x54:OR | n.a. | $27.5 \pm 10.3$ | 4 |
| 37ph | 0x50:OR | n.a. | $37.1 \pm 12.8$ | 6 |
| 52ph | 0x00:OR | 8-OR | $52.2 \pm 15.0$ | 8 |

The IF board houses an FPGA, which is used to control voltages and to configure and read out the DPCs of the sensor tile as well as a temperature sensor located on the back side of the latter. The main voltage lines controlled by the IF board are the $V_{util}$ and $V_{bias}$ lines; $V_{bias}$ sets the operating voltage of the SPADs. The SPADs should be operated at an excess voltage (overvoltage, $V_{ov}$) in a range of about $V_{ov} = 2\,V$–$3.3\,V$ above the measured breakdown voltage ($V_{bd}$) (manufacturer default: $V_{ov} = 3\,V$). $V_{util}$ is used to actively quench and recharge the SPADs and has to be larger than $V_{ov}$ to ensure the stable quenching of the SPADs. An additional safety margin of at least $0.2\,V$ should also be taken into account (Philips Digital Photon Counting 2014). The employed low-dropout regulator for $V_{util}$ introduces a 100 mA current limitation on the $I_{util}$ line. In particular, trig 1 causes a high load on the $I_{util}$ line due to the many dark-noise-induced triggers—especially at high temperatures—and may be influenced by the current limitation.

Six detector stacks are mounted on a singles detection module (SDM) in a $2 \times 3$ arrangement with a pitch of 33.3 mm (figure 1(b)). SDMs are equipped with an optical gigabit Ethernet (GbE) interface, cooled using a liquid cooling system with a process thermostat (Lauda Integral XT 150, Germany), and flooded with dry air. Liquid cooling pipes run between the IF board and the sensor tile, and the SDMs are shielded with a light- and RF-tight housing made from a 0.8 mm thick carbon fiber composite (Dueppenbecker *et al* 2012a).

The insert is composed of ten SDMs mounted on a gantry, and a synchronization unit distributes the reference clock and trigger signals. This results in a PET system with an inner diameter (distance of opposing inner crystal surfaces) of 209.6 mm and an axial field of view (FOV) of 97 mm (figure 2(a)). The gantry is able to house radio-frequency transmit and receive (RF Tx/Rx) coils with a high gamma transparency and different inner bore sizes (Weissler *et al* 2015), which are only installed as a passive component in this study. We mainly used a small RF Tx/Rx coil with an inner diameter of 46 mm, and to measure the rabbit-sized phantom, we installed a larger RF Tx/Rx coil with an inner diameter of 160 mm. The insert, the synchronization unit and a power supply are mounted on a trolley allowing easy installation in a Philips Achieva MRI system (figure 2(b)). All SDMs are linked via the GbE interfaces to a data acquisition and processing server (DAPS) which is connected to a control computer running monitoring and control software (Gebhardt *et al* 2012). The DAPS can be used to store the raw DPC sensor data for offline analysis, or it can process the data to that of the coincident list-mode in real time (Goldschmidt *et al* 2013, 2016). The control and status data are routed from the control PC to the system, and vice versa.

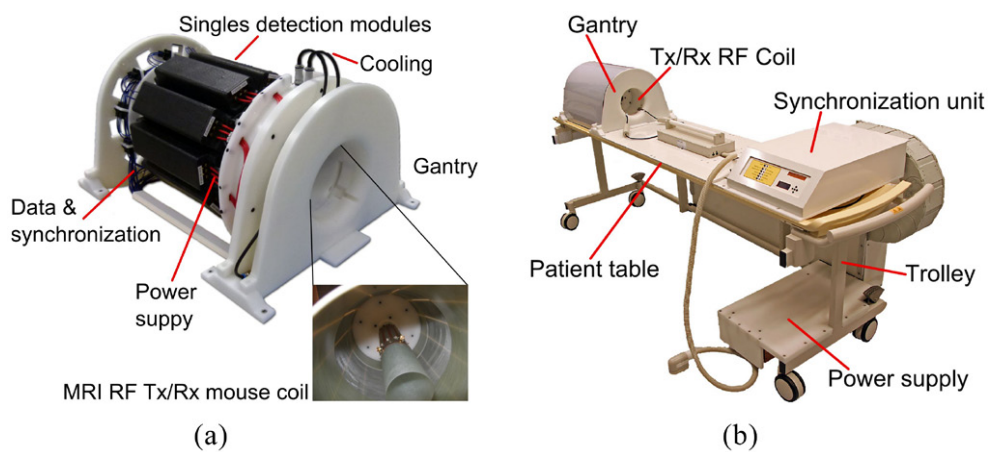

**Figure 2.** (a) The gantry of the PET insert holds ten SDMs (© [2012] IEEE. Reprinted, with permission, from Weissler *et al* (2012)). (b) The gantry, a synchronization unit and a power supply are mounted on a trolley allowing easy insertion into an MRI scanner. Only a single data link, a power cable and the cooling tubes have to be connected.

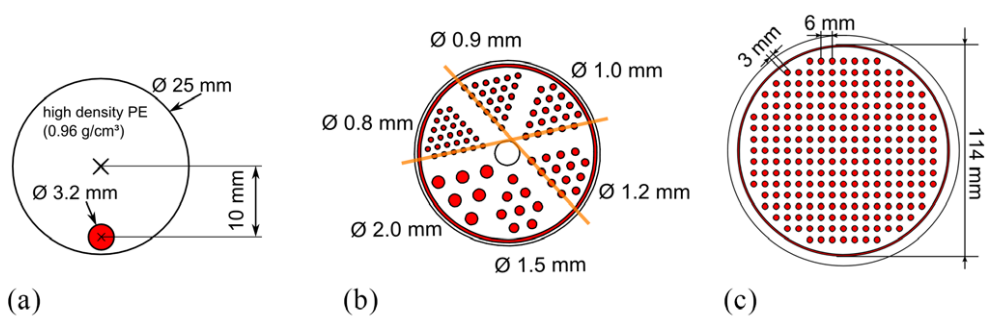

**Figure 3.** (a) A slice through the mouse-sized scatter phantom; the scatter phantom has an axial extent of 70 mm. (b) Hot-rod phantom used to investigate the spatial resolution; the structured region has a diameter of 28 mm, and the evaluated profiles are marked in orange. (c) Rabbit-sized phantom used to investigate the benefit of TOF information for image reconstruction; the structured part of the phantoms (b) and (c), for which the slices are shown, have an axial extent of 20 mm, and the FDG-filled areas are marked in red.

### 2.2. Test sources and phantoms

Five point-like $^{22}$Na sources with an active diameter of 0.25 mm and activity of 1.1 MBq–1.5 MBq enclosed in a cast acrylic cube with an edge length of 10 mm (NEMA cubes) were used for the performance studies under a large variety of different operating parameters at constant activity.

　　A mouse-sized scatter phantom (NEMA NU 4 standard) was used for the investigation of activity-dependent performance parameters including the count rate performance of the scanner (figure 3(a)). The phantom has an axial extent of 70 mm and a diameter of 25 mm with a hole to hold a tube with a radioactive tracer solution. This hole is placed 10 mm off center and is 3.2 mm in diameter. The tube fits the hole and has a wall thickness of 0.5 mm. The activity was distributed over the full 70 mm of the phantom opposed to the distribution defined in the NEMA NU 4 standard (60 mm).

　　For image-spatial-resolution studies, we used a mouse-sized hot-rod phantom with six regions and rods of 0.8 mm, 0.9 mm, 1.0 mm, 1.2 mm, 1.5 mm and 2.0 mm in diameter, a center-to-center spacing of twice the diameter and an axial extent of 20 mm (figure 3(b)).

　　To investigate the benefits of time-of-flight (TOF) image reconstruction, we used a rabbit-sized phantom with a diameter of 114 mm (figure 3(c)). The phantom is structured with rods with a diameter of 3 mm and an axial extent of 20 mm distributed on a Cartesian grid with a pitch of 6 mm.

　　All the phantoms were filled with a [$^{18}$F]Fluordeoxyglucose (FDG) solution.

## 3. Methods

In this work, we used the unprocessed raw DPC sensor data, and an offline analysis was performed using a calibration and processing framework which is described in detail in Schug *et al* (2015b). We employed a center-of-gravity (COG) method which defines a region of interest (ROI) around the pixel capturing most of the scintillation light (main pixel) of a single; the ROI includes the main pixel and the neighboring pixels. A COG method with automatic corner extrapolation (COG-ACE) was used throughout this work. It correctly handles up to two sets of DPC channels per crystal—allowing a single corner pixel to be missing—and yields a superior sensitivity compared to the COG method, which requires all neighboring channels (COG-FN). The DPC raw photon values were corrected for saturation (Schug *et al* 2015b). A correction of time stamps as a function of the light output of the scintillator (walk correction) was not applied.

### *3.1. Temperature measurements*

The cooling temperature $T_C$ was set and controlled by the process thermostat, and the operating temperature $T_{op}$ was measured with the temperature sensor on the back side of the sensor tiles. The system $T_{op}$ is reported as the mean and standard deviation of all the sensor tile $T_{op}$ readings.

### *3.2. Measurement of the breakdown voltage and bias voltage regulation*

By connecting the SPADs to the ground and measuring the current-voltage characteristic, $V_{bd}$ was determined for each sensor tile individually. This measurement was performed for different $T_C$ resulting in $T_{op} \approx 0\,°C$–$27\,°C$, and to determine the dependence of the DPC $V_{bd}$ on $T_{op}$, a linear regression on the measured data was performed.

　　For the following experiments, $V_{bias}$ was set to a constant value prior to the measurement and not dynamically adjusted during it. For one set of measurements with a different $T_C$, $V_{bias}$ was kept constant using the $V_{bd}$ value obtained at $T_C = 15\,°C$, which means that a change in $T_C$ leads to an effective change of $V_{ov}$. For the other set of measurements, however, $V_{bias}$ was adjusted before the start for the selected $T_C$, meaning that the effective $V_{ov}$ was kept constant as a function of $T_C$.

**Table 2.** Source positions for the measurements with the five NEMA cubes.

| NEMA cube | Constant $V_{bias}$ | | | | Constant $V_{ov}$ | | | |
|---|---|---|---|---|---|---|---|---|
| | $x$/mm | $y$/mm | $z$/mm | Activity/MBq | $x$/mm | $y$/mm | $z$/mm | Activity/MBq |
| 1 | −0.6 | −1.8 | −0.5 | 1.48 | −2.7 | −11.8 | −37.9 | 1.35 |
| 2 | −1.3 | −1.7 | 30.2 | 1.37 | −1.9 | −9.8 | −18.7 | 1.26 |
| 3 | −0.6 | −1.8 | −15.5 | 1.28 | −1.6 | −9.9 | −1.4 | 1.17 |
| 4 | −1.5 | −1.7 | 14.4 | 1.25 | −2.4 | −10.9 | 12.9 | 1.15 |
| 5 | −0.2 | −1.8 | −32.1 | 1.20 | −2.3 | −11.8 | 29.7 | 1.10 |

*3.3. Basic PET performance measurements*

We investigated the influence of the $V_{ov}$, $T_C$, trig, val, activity and energy window on the basic PET performance parameters: energy resolution ($\Delta E/E$), coincidence resolution time (CRT) and sensitivity. It is not feasible to scan all parameter combinations of the multi-dimensional space, therefore, we constructed several testing scenarios around conservatively chosen benchmark points and varied the parameters along some axes of the parameter space. For all measurements we inhibited 20% of the noisiest SPADs per pixel, used a validation length of 40 ns, an integration length of 165 ns and a fixed cluster window of 40 ns. We used two different energy windows: the narrow energy window (NE) ranges from 411 keV–561 keV to reject scattered gammas and discard high-light-output and pile-up events; the wide energy window (WE) was set to 250 keV–625 keV to increase the sensitivity compared to the NE. Singles were filtered before performing a coincidence search with a sliding coincidence window (CW), and if not mentioned otherwise, we used CW = 400 ps, 550 ps, 650 ps and 1500 ps for trig 1–4, respectively, which corresponds to approximately ±3σ CRT.

Coincidences with more than two singles were discarded, and a minimal distance of four detector stacks between the two singles in tangential direction was required. For the rabbit-sized phantom, this was reduced to three detector stacks to increase the FOV.

*3.3.1. $^{22}$Na point sources.* Five NEMA cubes were distributed in the FOV along the *z*-axis (axial), and the position and activity of the point sources are listed in table 2. This allows the CRT to be determined very precisely, as each line of response (LOR) can be unambiguously assigned to one of the point sources.

For the measurement series with constant $V_{bias}$, we chose $V_{ov}$ = 2.5 V, $T_C$ = 15 °C, trig 2 and val 17ph as our default benchmark point. Starting from this benchmark point, we investigated $V_{ov}$ = 2.5 V, 2.8 V, 2.9 V and 3.0 V and $T_C$ = −5 °C, 5 °C, 15 °C and 20 °C, trig 1–4 and val 17ph, 28ph and 52ph. Additionally, we measured at $T_C$ = −5 °C $V_{ov}$ = 2.5 V, 2.8 V and 2.9 V. We used the NE and WE to process the data, all performed measurements and their settings are listed in the supplementary table S1 (stacks.iop.org/PMB/61/2851/mmedia).

The benchmark point for the measurement with constant $V_{ov}$ = 2.5 V was chosen as val 17ph. We investigated $T_C$ = −5 °C, 5 °C, 10 °C, 15 °C and 20 °C for trig 1, 3 and 4. Trig 2 was omitted as it yields similar results to trig 3, as learned from the first measurement series; all measurements and their settings are listed in the supplementary table S2 (stacks.iop.org/PMB/61/2851/mmedia).

*3.3.2. FDG mouse-sized scatter phantom.* We used the mouse-sized scatter phantom filled with FDG (figure 3(a)) and started measuring at an activity of ∼110 MBq down to ∼0.5 kBq. We started with an acquisition time of approximately 30 s and increased it to about 6 min to partially compensate for declining activity. We used $T_C = 0\,°C$, $V_{ov} = 2.5\,V$ and programmed the DPCs with combinations of trig 1, 2 and 3 and val `17ph`, `28ph`, `37ph` and `52ph` during the decay of the FDG. All measurements and their settings are listed in the supplementary table S3 (stacks.iop.org/PMB/61/2851/mmedia).

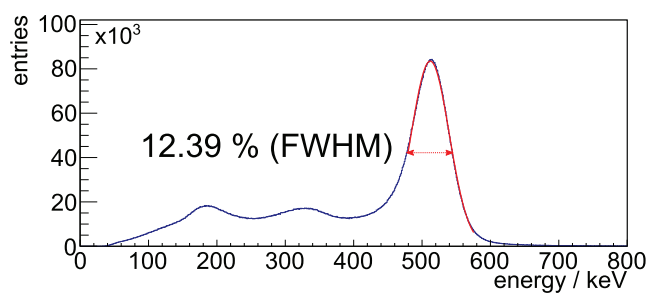

**Figure 4.** The exemplary and unfiltered system energy spectrum of coincident singles applying the COG-ACE algorithm requiring a minimum of 100 photons from a measurement with the FDG-filled mouse-sized scatter phantom and an activity of 8.41 MBq. The Gaussian fitted to the spectrum to evaluate the $\Delta E/E$ is plotted in the used fit range.

*3.3.3. Result extraction and computation.* $\Delta E/E$ was determined using the energy spectrum of coincident singles, and a Gaussian was iteratively fitted to match a fit range of $-0.5$ FWHM to $+1.0$ FWHM from the mean. No background removal or modeling was performed (see result plot figure 4).

To show the energy spectrum of the scanner over a larger range for one exemplary measurement of the mouse-sized scatter phantom, we applied a very small lower threshold of only 100 measured photons on the channels used for energy calculation on coincident singles (Schug *et al* 2015b).

The CRT was calculated by evaluating the FWHM of the measured timing difference spectrum between the coincident singles corrected for known source positions and either of the point sources or the line source. A Gaussian was iteratively fitted to match a fit range of $-0.5$ FWHM to $+0.5$ FWHM from the mean, and no background removal or modeling was performed, as the random fraction is small over the investigated parameter space (see results figures 8(c) and (d)).

For measurements using the $^{22}$Na point sources, the sensitivity was evaluated using the prompts rate without correcting for randoms and scatter. The random rate is expected to be below 1% at the given activity of the point sources (see results figures 8(c) and (d)); scatter in the point source material is expected to have a small effect and only relative comparisons are performed. The branching ratio of the $^{22}$Na $\beta^+$ decay of 0.906 was accounted for, and the sensitivity as a function of activity was measured with the mouse-sized scatter phantom corrected for scatter and randoms. The randoms rate was estimated based on the singles rate per crystal and was corrected for the prompts rate (Oliver and Rafecas 2012), and scatter estimation was performed based on the NEMA NU 4 standard using a single sinogram (projection of all LORS on a single transaxial plane) for the whole scanner. The prompts rate, randoms rate, scatter rate and NECR were evaluated in the corridor defined in the NEMA NU 4 standard.

### 3.4. Imaging experiments

All imaging experiments were conducted using the NE to suppress object scatter, and trig 1 and trig 3 were employed, and if not stated otherwise, the same parameters as for the basic

PET performance measurements were used. Reconstructions were performed for the whole datasets of the two measurements and, for comparison, they were trimmed once for the same number of tracer decays and once for the same number of recorded coincidences. This allowed comparisons which took the respective sensitivity into account as well as neglecting it and using the same statistics.

Image reconstructions were performed using an ordered subset expectation maximization (OSEM) (Hudson and Larkin 1994) 3D reconstruction (Salomon *et al* 2011, 2012). After each subset of an iteration, 3D Gaussian smoothing was performed, and we used self-normalization but did not correct for scatter and attenuation. Furthermore, the reconstructed activity distribution was linearly normalized using the same region of interest per phantom. The 3D data was projected on a transaxial plane using a defined extent in the axial direction (slice thickness), in order to reduce the required measurement time and acquire enough coincidences for image reconstruction showing the resolution of the scanner. Specific parameters for the reconstructions are stated for the respective phantom measurement, and convergence was checked and reached for all the reconstructed images.

All of the above-mentioned datasets were reconstructed with and without TOF information, and we calculated the absolute difference between the two reconstructions for each dataset. The absolute-difference images were multiplied by a factor of 5 in order to better visualize the difference using the same grayscale as used for the reconstructed data sets.

*3.4.1. FDG mouse-sized hot-rod phantom.* For the mouse-sized hot-rod phantom (figure 3(b)), the measurement parameters were chosen as $T_C = -5\,°C$, $V_{ov} = 2.5$ V and val `28ph`. The trig 3 measurement was started with an activity of 9.0 MBq for a measurement time of 762 s, and approximately 8 min later, the phantom was measured at an activity of 7.9 MBq with trig 1 for 1160 s.

For the image reconstructions, we used a voxel pitch of 0.25 mm, a matrix size of $250 \times 250 \times 387$, 16 iterations, 32 subsets, and a slice thickness of 20 mm. We also applied a Gauss filter of approximately 0.17 mm (FWHM) after each subset. We extracted two profiles through these slices: one profile going through the rods with a diameter of 0.9 mm and 1.2 mm, and a second profile going through the rods with a diameter of 0.8 mm and 1.0 mm (figure 3(b)). We calculated the peak-to-valley values for each peak using the profiles, and the peak height was divided by the mean height of the two adjacent valleys or, for the first and last peak, the height of the single adjacent valley.

*3.4.2. FDG rabbit-sized phantom.* For the rabbit-sized phantom (figure 3(c)), we used $T_C = -5\,°C$, $V_{ov} = 2.5$ V and val `28ph`, and a CW of 1.5 ns was used for both trigs to account for the diameter of the activity distribution.

For the first experiment using this phantom, the axes of the Cartesian grid on which the rods were located were aligned with the axes of the transversal plane of the scanner (see section 5.5 and figure 13). The trig 3 measurement was started with an activity of 3.8 MBq for a measurement time of 1542 s, and approximately 14 min later, the phantom was measured at an activity of 3.0 MBq using trig 1 for 1018 s.

For a second experiment, the phantom was rotated around the axial axis by 9° with respect to the first experiment to break the alignment of the phantom axes with the scanner gaps (see discussion section 5.5). For the trig 1 measurement, an activity of 13.8 MBq was used and we measured it for a duration of 924 s; trig 3 was measured approximately 8 min later with an activity of 11.9 MBq for 708 s.

For the image reconstructions, we used a voxel pitch of 1 mm, a matrix size of 200 × 200 × 97, 16 iterations, 8 subsets and a slice thickness of 10 mm. We also applied a Gauss filtering after each subset of 0.7 mm (FWHM).

The signal-to-background ratio was computed with a signal region defined as the tracer distribution (figure 3(c)) with an extension of the region by 1 mm in all directions to account for the resolution of the scanner and the voxel size of the reconstruction. The background region is defined as the image region not belonging to the signal region. To calculate the total activity sum for both regions, the reconstructed images were sampled with a resolution of 0.05 mm (the signal regions superimposed on the reconstructed images are shown in the supplementary figures S2 and S3 (stacks.iop.org/PMB/61/2851/mmedia)). This evaluation allows the difference between trigger settings and the benefit of using TOF information during reconstruction to be quantified. A peak-to-valley calculation is not practicable as the inhomogeneous image quality and the large structures of the phantom compared to the scanner's spatial resolution cause very different results depending on where the line profile is chosen.

## 4. Results

Detailed results of all measurements using the point sources and the mouse-sized scatter phantom are listed in the supplement (tables S1–S3), and we extracted graphs showing the behavior of the performance parameter as a function of $V_{ov}$, system $T_{op}$ and activity. The difference of $T_C$ and the system's $T_{op}$ was approximately 5 °C–10 °C under operation, mainly depending on $T_C$, trig and activity (for all measurements, both values are reported in the supplement); improvements and degradations are reported as relative changes.

### 4.1. Measurement of the breakdown voltage

The mean and standard deviation of $V_{bd}$ for all sensor tiles was determined as 23.02 ± 0.12 V at $T_{op} = 15$ °C and its dependence on $T_{op}$ as 17.1 ± 1.0 mV K$^{-1}$ (supplementary figure S1 (stacks.iop.org/PMB/61/2851/mmedia)). The $T_{op}$ of one detector stack could not be read out due to a broken sensor and was omitted for the evaluation, nonetheless, $V_{bd}$ was saved for the applied values of $T_C$ for this detector stack as well.

### 4.2. Possible operating parameters

Due to the current limitations on the $I_{util}$ line, using trig 1 was limited to low temperatures and low activity. $T_C = 5$ °C ($T_{op} = 13.77 \pm 1.35$ °C) was the highest of the tested temperatures which allowed a stable operation of trig 1 using $V_{ov} = 2.5$ V. The highest possible activity was 36.74 MBq, and the other trig could be used without restrictions for all tested operation conditions.

### 4.3. System energy resolution

The exemplary energy spectrum for coincident singles obtained from a measurement with the FDG-filled scatter phantom for an activity of 8.41 MBq with $\Delta E/E = 12.39\%$ is shown in figure 4.

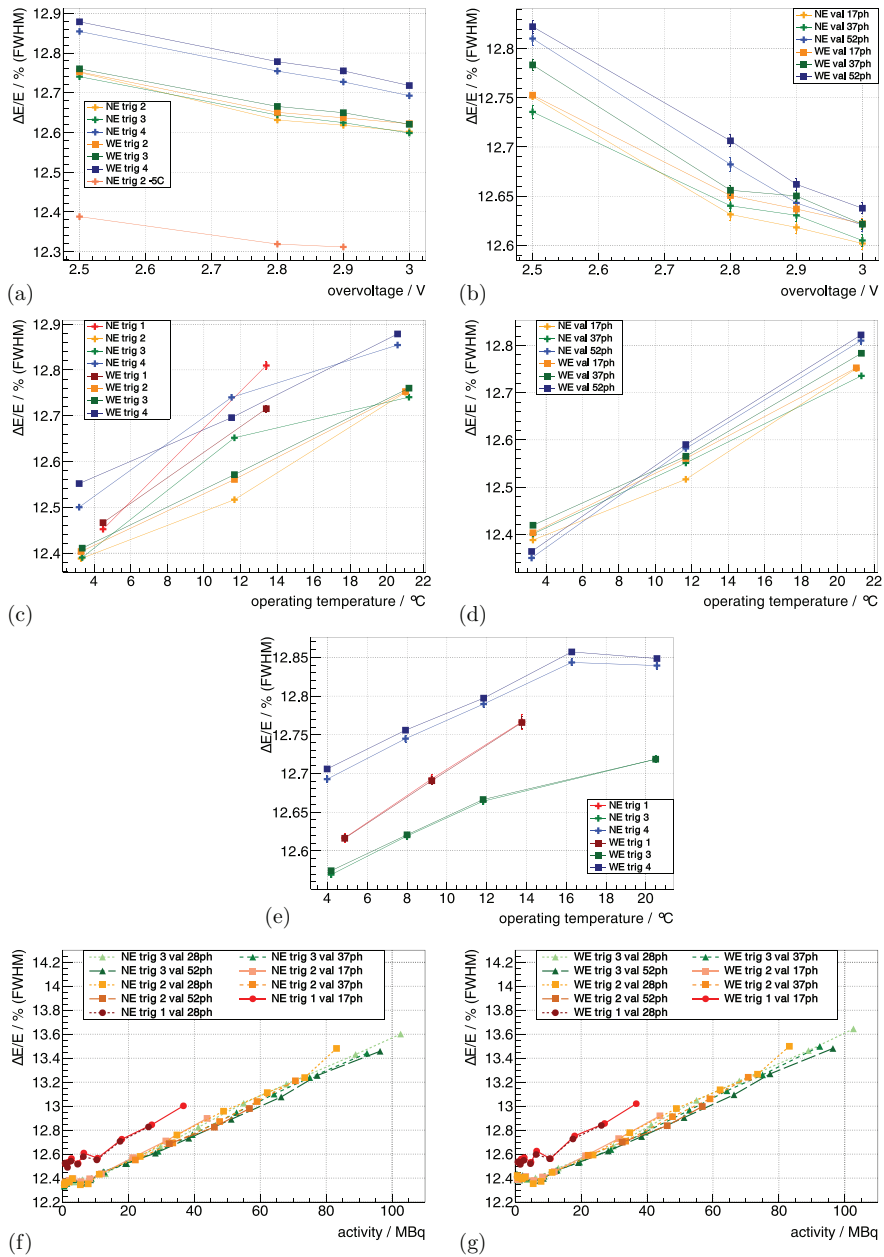

**Figure 5.** $\Delta E/E$ of the whole insert measured with the $^{22}$Na point sources is shown as a function of $V_{ov}$ in (a) for different trig and in (b) for different val. It is then shown as a function of $T_{op}$ for a constant $V_{bias}$ in (c) for different trig and in (d) for different val. For a constant $V_{ov}$, it is shown as a function of $T_{op}$ in (e) for different trig. Measured with the FDG-filled scatter phantom, it is shown in (f) and (g) as a function of activity for different trig and val. (a) $T_C = 15\,°C$, val | 7ph. (b) $T_C = 15\,°C$, trig 2. (c) const $V_{bias}$, $V_{ov}(T_C = 15\,°C) = 2.5\,V$, val | 7ph. (d) const $V_{bias}$, $V_{ov}(T_C = 15\,°C) = 2.5\,V$, trig 2. (e) const $V_{ov}(T_C) = 2.5\,V$, val | 7ph. (f) $T_C = 0\,°C$, $V_{ov} = 2.5\,V$. (g) $T_C = 0\,°C$, $V_{ov} = 2.5\,V$.

Using the $^{22}$Na sources at the default benchmark point (see methods section 3.3.1), $\Delta E/E$ for trig 2, trig 3 and trig 4 was measured as 12.75%, 12.75% and 12.85%. Increasing the $V_{ov}$ from 2.5 V to 3 V improved $\Delta E/E$ relatively by 1%–2% (figures 5(a) and (b)).

For constant $V_{bias}$, decreasing $T_C$ from 15 °C down to −5 °C ($T_{op}$ from ∼21 °C down to ∼3 °C) improved $\Delta E/E$ relatively by ∼2.5%–3.5% (figures 5(c) and (d)). In the evaluated $T_{op}$ range, trig 1 showed a dependence that was approximately twofold stronger. For constant $V_{ov}$, the dependence of $\Delta E/E$ on $T_{op}$ was smaller and was measured as a relative change of ∼1% for a comparable change in $T_{op}$ (figure 5(e)). $\Delta E/E$ for trig 4 was relatively ∼1% worse compared to trig 2 and 3 for all operating conditions.

$\Delta E/E$ determined with the measurement of the mouse-sized scatter phantom as a function of activity is shown in figures 5(f) and (g). It degraded relatively by ∼9% from low activity to 100 MBq for all trig; trig 1 showed an $\Delta E/E$ which was relatively ∼1% worse compared to the other trig.

### 4.4. Coincidence resolution time

Using the $^{22}$Na sources at the default benchmark point (see methods section 3.3.1), the system's CRT using the NE for trig 2, trig 3 and trig 4 was measured as 450 ps, 562 ps and 1.3 ns, respectively. Increasing $V_{ov}$ from 2.5 V to 3 V improved the CRT up to ∼9% (figures 6(a) and (b)), and higher val showed a stronger improvement.

For constant $V_{bias}$, decreasing $T_C$ from 15 °C down to −5 °C ($T_{op}$ from ∼21 °C down to ∼3 °C) improved the CRT values up to 15%–20%. When operating at a $T_{op}$ of 4.48 ± 1.45 °C and 13.41 ± 1.32 °C, trig 1 delivered a CRT of 258 ps and 272 ps (figures 6(c) and (d)). For constant $V_{ov}$, the dependence of CRT on $T_{op}$ was mainly eliminated (figure 6(e)). The CRT values measured at $T_C = 15$ °C ($T_{op} \approx 20.5$ °C) were lower than the ones obtained at all other $T_{op}$. The trig 1 showed a degradation of the CRT value when measuring $T_{op} = 13.77 ± 1.35$ °C compared to the other $T_{op}$, and the CRT of the WE was 15%–20% worse compared to the NE.

Using the mouse-sized scatter phantom, the CRT as a function of activity is shown in figures 6(f) and (g). The CRT degraded linearly by about 3%–5% from small activity to 100 MBq, and trig 1 showed a dependence more than twofold stronger.

### 4.5. Sensitivity

The prompts rate for the five $^{22}$Na point sources at the default benchmark point (see methods section 3.3.1) was measured as 60 kcps using the NE and 124 kcps for the WE, which results in a sensitivity of 1.0% and 2.1% for the given distribution of point sources. Increasing the $V_{ov}$ from 2.5 V to 3 V only improved the sensitivity for trig 4— the other trig lost up to 10% sensitivity (figure 7(a)). For high val, increasing the $V_{ov}$ was beneficial, but low val, on the other hand, lost sensitivity (figure 7(b)).

For constant $V_{bias}$, decreasing $T_{op}$ improved the sensitivity for almost all operating parameters— only the sensitivity for trig 3 and 4 degraded when going from $T_{op} \approx 11.5$ °C to $T_{op} \approx 3.0$ °C and applying the NE (figures 7(c) and (d)). For constant $V_{ov}$, lower $T_{op}$ were beneficial for all trig figure 7(e). In particular, the sensitivity of trig 1 was strongly dependent on $T_{op}$, which showed a loss of ∼35% at $T_{op} = 4.86 ± 1.42$ °C up to ∼65% at $T_{op} = 13.77 ± 1.35$ °C compared to the higher trig.

Using the mouse-sized scatter phantom, the scatter fraction was measured to be ∼6% for the NE (figure 8(a)) and ∼16% for the WE (figure 8(b)), showing a slight increase as a function of activity. For the NE it was almost independent of the used trig and val, and for the WE, the maximal differences were within a relative band of ∼±15%.

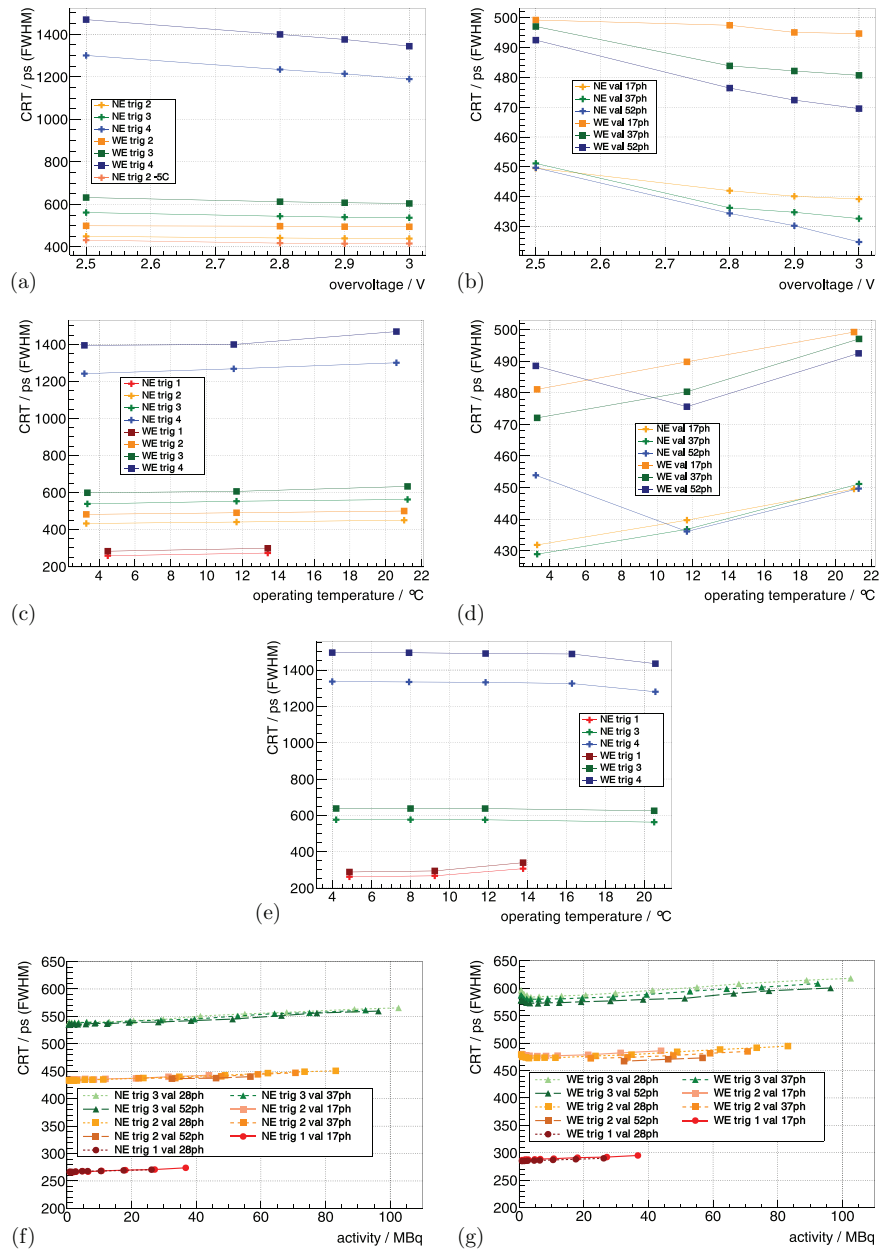

**Figure 6.** The CRT measured with the $^{22}$Na point sources is shown in (a) as a function of $V_{ov}$ for different trig and in (b) for different val. It is shown for a constant $V_{bias}$ as a function of $T_{op}$ in (c) for different trig and in (d) for different val. For a constant $V_{ov}$, it is shown as a function of $T_{op}$ in (e) for different trig. Measured with the FDG-filled scatter phantom, it is shown in (f) and (g) as a function of activity for different trig and val. (a) $T_C = 15\,°C$, val | 7ph. (b) $T_C = 15\,°C$, trig 2. (c) const $V_{bias}$, $V_{ov}(T_C = 15\,°C) = 2.5$ V, val | 7ph. (d) const $V_{bias}$, $V_{ov}(T_C = 15\,°C) = 2.5$ V, trig 2. (e) const $V_{ov}(T_C) = 2.5$ V, val | 7ph. (f) $T_C = 0\,°C$, $V_{ov} = 2.5$ V. (g) $T_C = 0\,°C$, $V_{ov} = 2.5$ V.

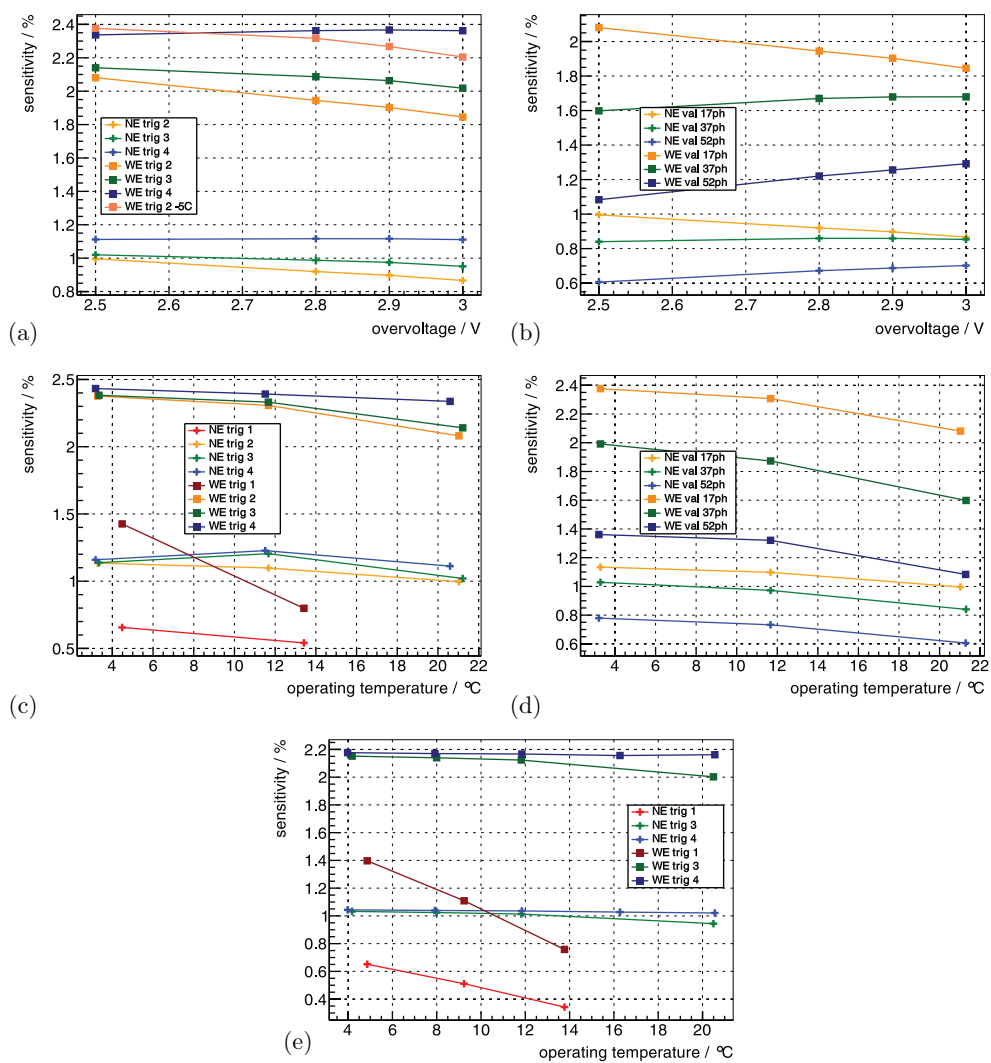

**Figure 7.** The sensitivity for the $^{22}$Na point sources (table 2) is shown as a function of $V_{ov}$ in (a) for different trig and in (b) for different val. It is shown for a constant $V_{bias}$ as a function of $T_C$ in (c) for different trig and in (d) for different val. For a constant $V_{ov}$, it is shown as a function of $T_C$ in (e) for different trig. (a) $T_C = 15\,°C$, val | 7ph. (b) $T_C = 15\,°C$, trig 2. (c) const $V_{bias}$, $V_{ov}(T_C = 15\,°C) = 2.5\,V$, val | 7ph. (d) const $V_{bias}$, $V_{ov}(T_C = 15\,°C) = 2.5\,V$, trig 2. (e) const $V_{ov}(T_C) = 2.5\,V$, val | 7ph.

The random fraction of the system was dependent on the used energy window and the CW applied (figures 8(c) and (d)). This can be roughly approximated with a linear dependence on the activity of ∼0.02%/MBq–0.05%/MBq.

The NECR curves of the mouse scatter phantom showed different peak NECRs dependent on the trig and val (figures 8(e) and (f)); the peak NECR was beyond 25 MBq for all parameters. Using WE, trig 2 and 3 with a val of at least `28ph`, we measured peak NECRs of about 280 kcps–320 kcps, and for higher trig and val the peak NECR shifted to activity of up to 50 MBq.

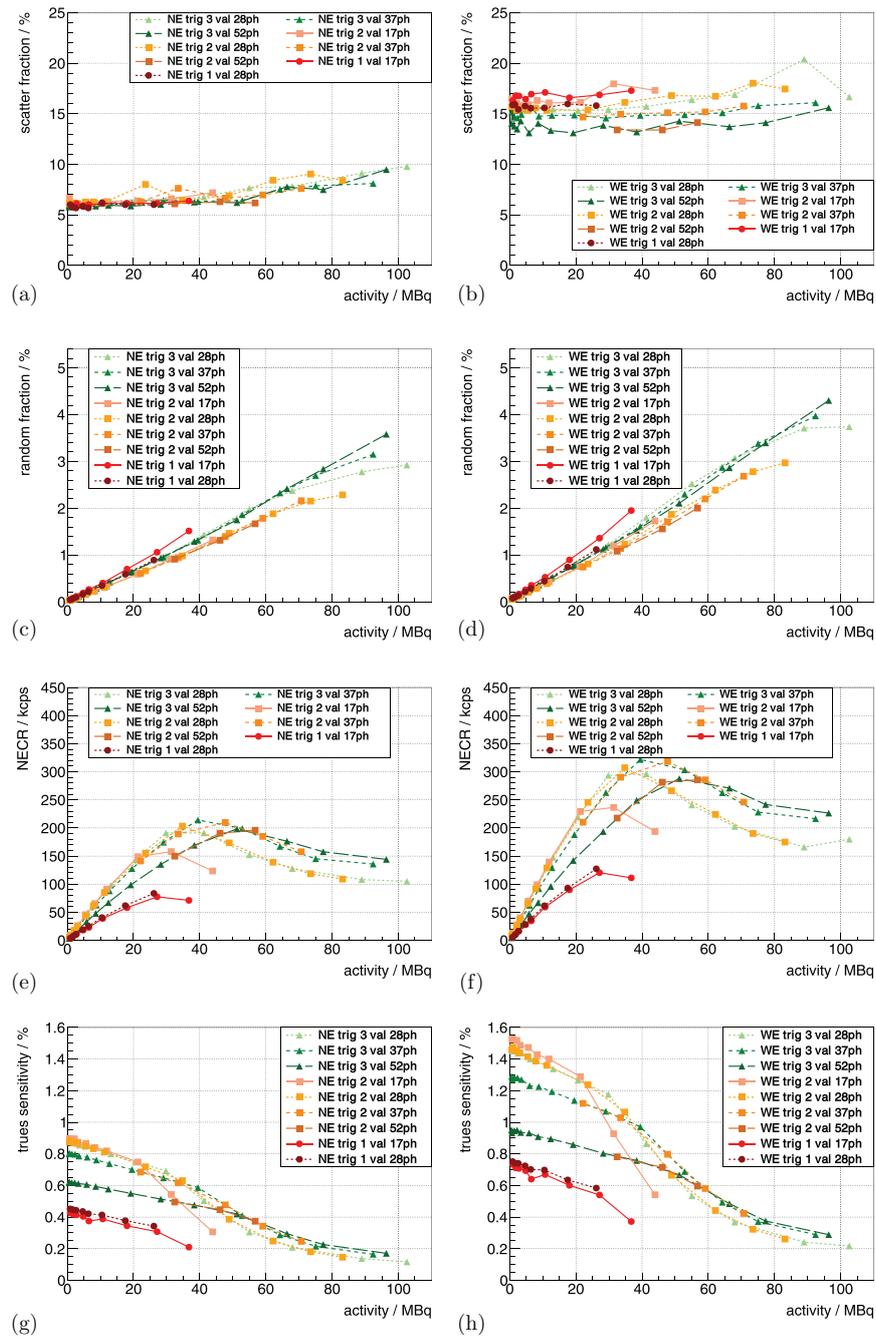

**Figure 8.** (a) and (b) show the scatter fraction curves for the mouse-sized scatter phantom for a constant $V_{\text{ov}} = 2.5$ V and $T_{\text{C}} = 0\,°\text{C}$ for different trig and val. The two bands for the different energy windows can clearly be distinguished. For the same measurement, (c) and (d) show the random fraction, (e) and (f) the NECR curves and (g) and (h) the trues sensitivity.

**Table 3.** Statistics for the hot-rod phantom measurement.

|     | Trig | Normalized to | Activity | System $T_{op}$ | Meas. time | FDG decays | Counts |
| --- | --- | --- | --- | --- | --- | --- | --- |
| (a) | 3 | All | 9.0 MBq | 3.52 ± 1.38 °C | 762 s | $6.48 \times 10^9$ | $66.59 \times 10^6$ |
| (b) | 1 | Decays | 7.9 MBq | 3.38 ± 1.29 °C | 875 s | $6.48 \times 10^9$ | $42.99 \times 10^6$ |
| (c) | 3 | Counts | 9.0 MBq | 3.52 ± 1.38 °C | 640 s | $5.49 \times 10^9$ | $56.25 \times 10^6$ |
| (d) | 1 | All | 7.9 MBq | 3.38 ± 1.29 °C | 1160 s | $8.41 \times 10^9$ | $56.25 \times 10^6$ |

The trues sensitivity curve for the mouse scatter phantom is shown in figures 8(g) and (h), and for low activity, an approximately linear decrease in trues sensitivity could be observed, followed by a stronger decline in sensitivity.

### 4.6. FDG mouse-sized hot-rod phantom

The results and statistics for the mouse-sized hot-rod phantom measurements are listed in table 3. The measurements normalized for the same amount of decays during the measurement time are shown for trig 3 and trig 1 in figures 9(a) and (b) and those normalized for the same number of coincidences in figures 9(c) and (d). None of the measurements showed a benefit of the TOF reconstruction.

The profile lines through the slices, as defined in figure 3, are shown for 0.8 mm and 1.0 mm rods in figure 10(a) and for 0.9 mm and 1.2 mm rods in figure 10(b). As the profiles showed a systematic dependency of peak-to-valley values on the position and did not differ significantly for the different measurement settings and reconstructions, we only state the combined mean peak-to-valley value for each rod size of all the shown profiles. The extracted mean peak-to-valley values and their standard deviations are $1.24 \pm 0.15$, $1.47 \pm 0.18, 1.87 \pm 0.19$ and $2.32 \pm 0.28$ for rod sizes of 0.8 mm, 0.9 mm, 1.0 mm and 1.2 mm, respectively.

### 4.7. FDG rabbit-sized phantom

For the measurement with the axes of the Cartesian grid of the phantom aligned with the axes of the transversal plane of the scanner, the results and statistics are listed in table 4. The measurements normalized for the same amount of decays during the measurement time are shown in figures 11(a) and (b) and those normalized for the same number of coincidences in figures 11(b) and (c). The untrimmed trig 3 measurement with threefold the number of counts compared to trig 1 is shown in figure 11(d).

For the measurement in the tilted position, the results and statistics are listed in table 5. The measurements normalized for the same amount of decays during the measurement time are shown in figures 12(a) and (b) and those normalized for the same number of coincidences are shown in figures 12(c) and (d).

For both phantom orientations, the trig 1 measurements (figures 11(b) and (d)) showed a benefit of TOF reconstruction.

The signal-to-background values (table 6) confirm the visual impression; for non-TOF reconstructions the higher sensitivity of trig 3 shows an advantage compared to trig 1. Including the TOF information improves the signal-to-background values for trig 1 reconstructions by about 21% and trig 3 reconstruction by about 7%, and, furthermore, the best signal-to-background values are obtained for trig 1 reconstructions with TOF information.

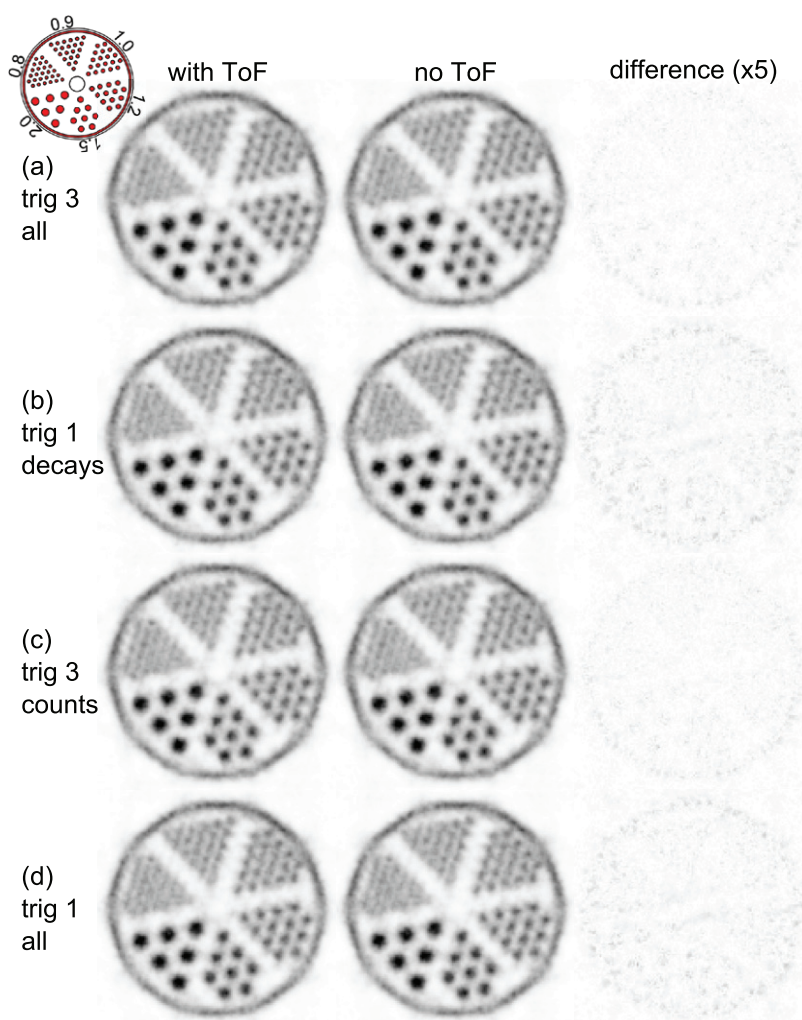

**Figure 9.** Transversal slices through the reconstructed hot-rod phantom. The voxel pitch is 0.25 mm, and the slice thickness is 20 mm. The complete datasets of trig 1 and trig 3 (all) are truncated for comparison: *trig 3 all* matches *trig 1 decays* in terms of the number of tracer decays and *trig 3 counts* matches *trig 1 all* in terms of the number of coincidences (table 3). The third column shows the absolute difference multiplied by a factor of 5 between the two reconstructions for each measurement.

## 5. Discussion

$\Delta E/E$ and CRT are comparable to the results shown with similar scintillators on the PDPC TEK platform (Dueppenbecker *et al* 2011, Degenhardt *et al* 2012). We thus conclude that the Hyperion-II$^D$ data acquisition platform provides a stable clocking and voltage environment at a system level. In other studies, we showed that our platform is capable of being operated simultaneously inside an MRI (Wehner *et al* 2014, Schug *et al* 2015a, Wehner *et al* 2015, Weissler *et al* 2015).

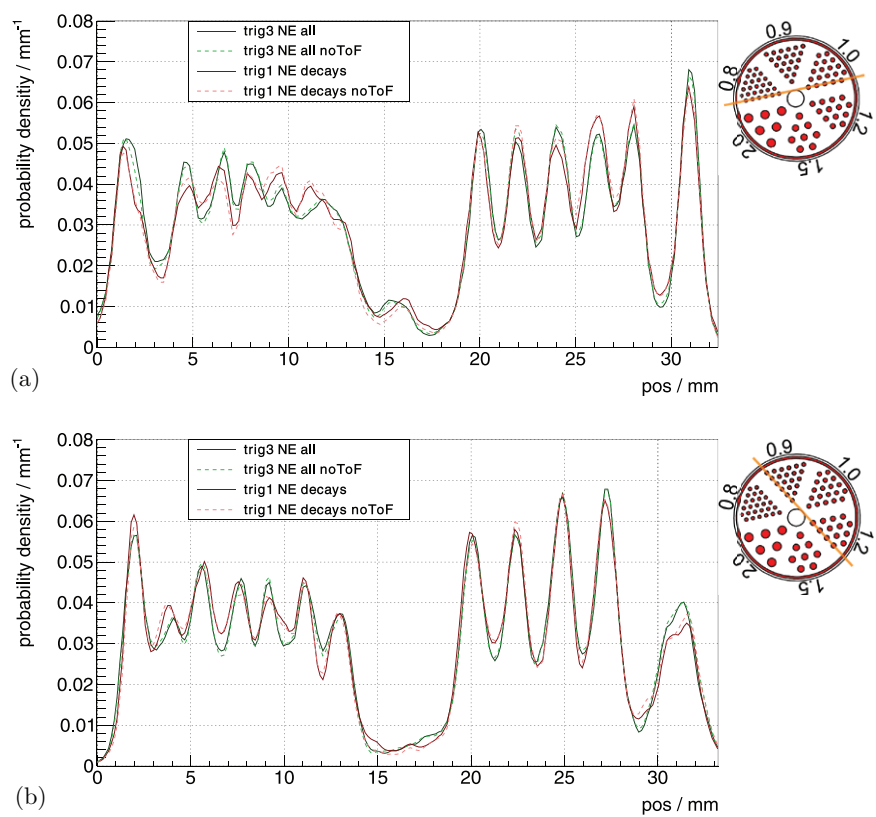

**Figure 10.** Profiles of the hot-rod phantom through the (a) 0.8 mm and 1.0 mm rods and (b) 0.9 mm and 1.2 mm rods.

**Table 4.** Statistics for the rabbit-sized phantom measurement.

|     | Trig | Normalized | Activity | System $T_{op}$ | Meas. time | FDG decays | Counts |
| --- | --- | --- | --- | --- | --- | --- | --- |
| (a) | 3 | Decays | 3.8 MBq | $3.85 \pm 1.14$ °C | 779 s | $2.79 \times 10^9$ | $15.88 \times 10^6$ |
| (b) | 1 | All | 3.0 MBq | $4.83 \pm 1.39$ °C | 1018 s | $2.79 \times 10^9$ | $9.90 \times 10^6$ |
| (c) | 3 | Counts | 3.8 MBq | $3.85 \pm 1.14$ °C | 478 s | $1.75 \times 10^9$ | $9.90 \times 10^6$ |
| (d) | 3 | All | 3.8 MBq | $3.85 \pm 1.14$ °C | 1542 s | $5.23 \times 10^9$ | $30.24 \times 10^6$ |

## 5.1. Possible operating parameters

The current limitation on the $I_{util}$ line introduces limitations on the possible operating parameters. For example, high operating temperatures cause high DCRs, and in combination with low trig, this produces noise triggers which increase the power consumption and dead time of the DPCs. $T_C = 5$ °C ($T_{op} = 13.77 \pm 1.35$ °C) was the highest temperature applied using trig 1 which allowed stable operation. Furthermore, trig 1 was only applicable up to activities of 36.74 MBq due to the current limitation on the $I_{util}$ line. Even without these restrictions, trig 1 should not be used at $T_{op} \gtrsim 5$ °C–10 °C as this introduces significant dead time. The higher trig, however, are more robust to high temperatures but still benefit from low

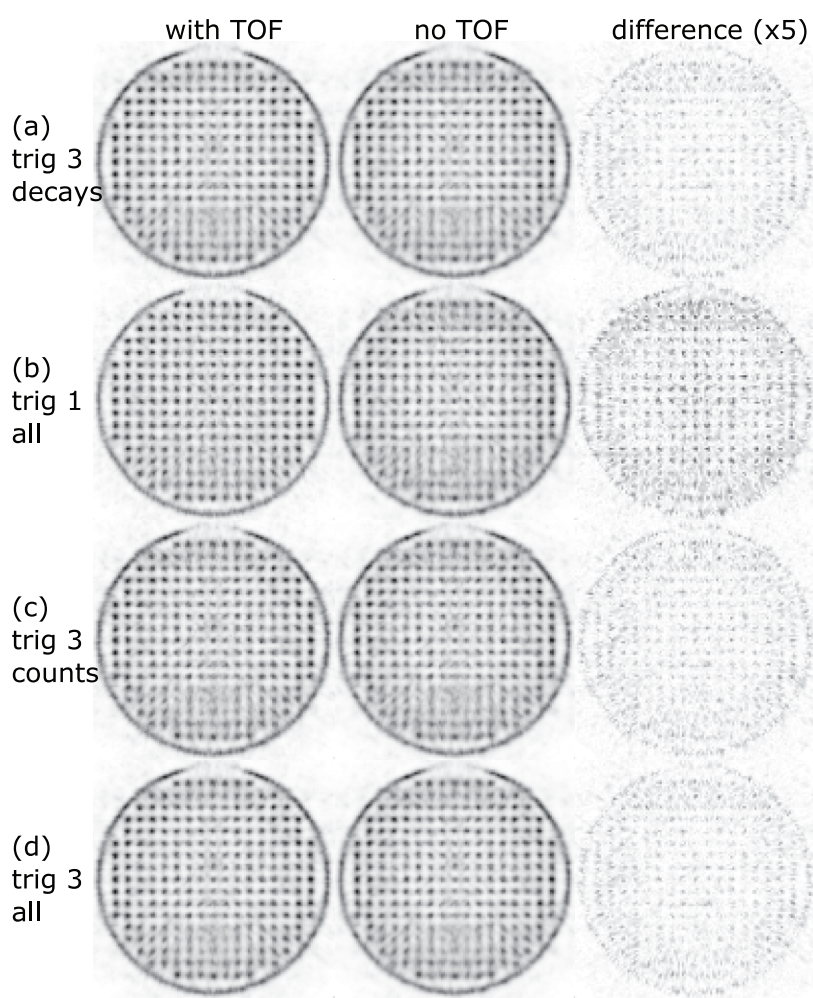

**Figure 11.** Transversal slices through the reconstructed rabbit-sized phantom. The voxel pitch is 1 mm, and the slice thickness is 10 mm. The complete datasets of trig 1 and trig 3 (all) are truncated for comparison: *trig 3 decays* matches *trig 1 all* in terms of the number of tracer decays and *trig 3 counts* matches *trig 1 all* in terms of the number of coincidences (table 4). The third column shows the absolute difference multiplied by a factor of 5 between the two reconstructions for each measurement.

**Table 5.** Statistics for the rabbit-sized phantom measurement in the tilted position.

|  | Trig | Normalized | Activity | System $T_{op}$ | Meas. time | FDG decays | Counts |
|---|---|---|---|---|---|---|---|
| (a) | 3 | All | 11.9 MBq | 3.49 ± 1.15 °C | 708 s | $7.99 \times 10^9$ | $43.47 \times 10^6$ |
| (b) | 1 | Decays | 13.8 MBq | 3.66 ± 1.26 °C | 606 s | $7.99 \times 10^9$ | $26.70 \times 10^6$ |
| (c) | 3 | Counts | 11.9 MBq | 3.49 ± 1.15 °C | 652 s | $7.38 \times 10^9$ | $40.10 \times 10^6$ |
| (d) | 1 | All | 13.8 MBq | 3.66 ± 1.26 °C | 924 s | $11.89 \times 10^9$ | $40.10 \times 10^6$ |

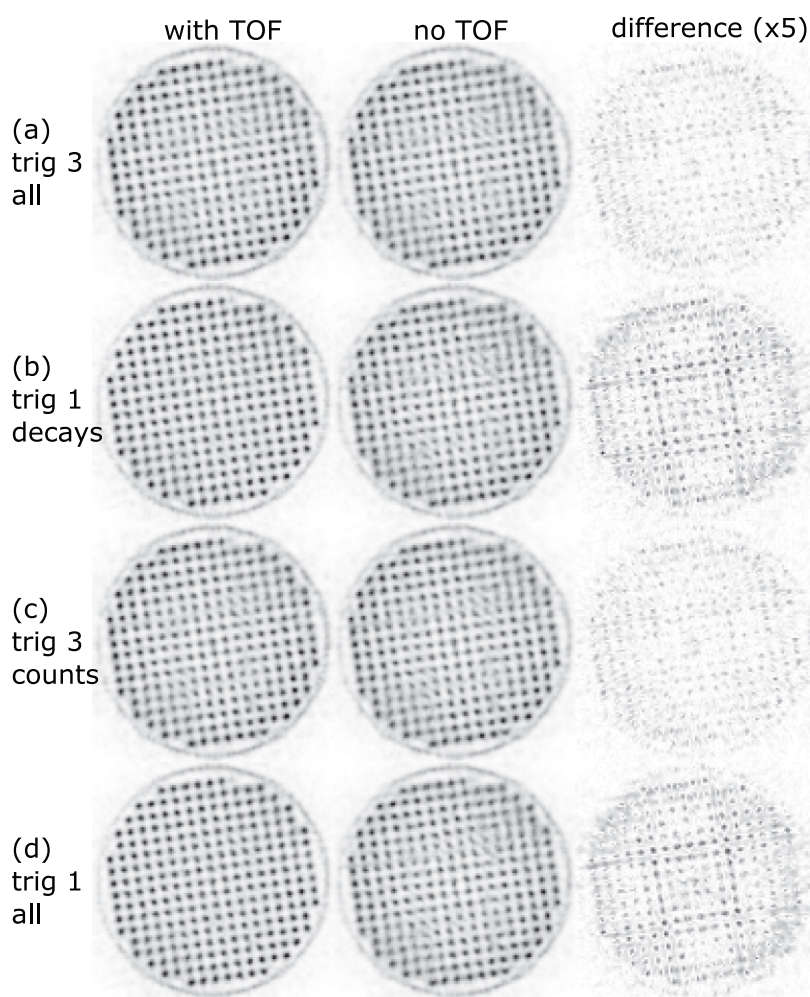

**Figure 12.** Transversal slices through the reconstructed rabbit-sized phantom measured in the tilted position. The voxel pitch is 1 mm, and the slice thickness is 10 mm. The complete datasets of trig 1 and trig 3 (all) are truncated for comparison: *trig 3 all* matches *trig 1 decays* in terms of the number of tracer decays and *trig 3 counts* matches *trig 1 all* in terms of the number of coincidences (table 5). The third column shows the absolute difference multiplied by a factor of 5 between the two reconstructions for each measurement.

temperatures—especially in terms of sensitivity. Limits of the platform due to the saturation of the GbE interfaces are discussed in section 5.4.

### 5.2. Energy resolution

The $\Delta E/E$ of about ~12.7% was very stable for all combinations of parameters applied. At this level, the scanner outperforms the $\Delta E/E$ shown in other preclinical systems (Inveon: 14.6% (Bao *et al* 2009), PET Component of the NanoPET/CT: 19% (Szanda *et al* 2011), LabPET: 25% (Bergeron *et al* 2014)). A good energy resolution allows to discriminate the photo-peak from the scattered events. The temperature dependency was smaller for constant

**Table 6.** Results of the signal-to-background (S2B) evaluation of the image reconstructions of the rabbit-sized phantom.

|  | Trig | Normalized | S2B (with TOF) | S2B (no TOF) | TOF change |
|---|---|---|---|---|---|
| Un-tilted | 3 | Decays | 2.41 | 2.25 | 7.10% |
|  | 1 | All | 2.69 | 2.23 | 20.33% |
|  | 3 | Counts | 2.37 | 2.21 | 6.85% |
|  | 3 | All | 2.41 | 2.25 | 7.15% |
| Tilted | 3 | All | 2.31 | 2.15 | 7.68% |
|  | 1 | Decays | 2.59 | 2.13 | 21.82% |
|  | 3 | Counts | 2.31 | 2.15 | 7.53% |
|  | 1 | All | 2.61 | 2.14 | 22.14% |

Note: the signal-to-background values are reported for each of the reconstructions shown in figures 11 and 12. In the last column, the relative change of the signal-to-background value using the TOF information compared to the non-TOF case is reported.

$V_{ov}$, as expected. The slight degradation towards higher activity can most likely be attributed to pile-up effects.

### 5.3. Coincidence resolution time

The CRT for trig 1 is the best that has been shown so far for a system with a high-resolution pixelated scintillator configuration. It is at the same level as the CRT obtained with $4 \times 4 \times 22\,mm^3$ one-to-one coupled LYSO crystals on a demonstrator built by PDPC (Degenhardt *et al* 2012). Using the Hyperion-II$^D$ platform with $4 \times 4 \times 10\,mm^3$ one-to-one coupled LYSO crystals, we recently achieved a CRT of 215 ps (Schug *et al* 2015c), which is only a difference on the order of 20% compared to the high-resolution scintillator configuration presented here. Furthermore, the CRT values obtained with trig 2 and 3 still outperform, to our knowledge, the commercial preclinical systems available today (Inveon: 1.22 ns (Lenox *et al* 2006), PET Component of the NanoPET/CT: 1.5 ns–3.2 ns (Szanda *et al* 2011), LabPET: ∼9 ns (Bergeron *et al* 2014)).

Generally, the CRT improved with increasing $V_{ov}$, and the effect was higher for high val as the noise-induced DPC hits have a smaller statistical weight when determining the final CRT. The CRT for the WE is expected to be worse, as no energy-dependent walk correction for the time stamps was performed.

One measurement with constant $V_{bias}$ at $T_C = -5\,°C$ ($T_{op} = 3.21 \pm 1.15\,°C$), $V_{ov} = 2.5$ V, val 52ph and trig 2 showed a deviation from the expected behavior in terms of CRT (see figure 6(d) and table S1 measurement 11 and 12). Our only explanation is that the configuration of the system was not applied as expected. Furthermore, the measurement with constant $V_{ov}$ at $T_C = 15\,°C$ ($T_{op} \approx 20.5\,°C$) showed a better CRT performance than the ones obtained at all other temperatures (see table S2 measurements 21–24). We can only explain this behavior due to a faulty determination of $V_{bd}$ at this temperature leading to a different effective $V_{ov}$ compared to the other measurements.

### 5.4. Sensitivity

A higher $V_{ov}$ increases the PDE of the DPCs as well as the DCR and cross talk. Therefore, increasing the $V_{ov}$ for low val causes more dark noise events to be validated and thus increases the dead time, reducing the system's sensitivity. If a high val is applied, not all the DPCs of low-energy events may be validated and a higher $V_{ov}$ increases the probability of reaching the validation threshold, thus increasing the system's sensitivity.

The random fraction measured with the mouse-sized scatter phantom is low compared to commercially available systems (Goertzen *et al* 2012), which is due to the superior CRT performance and the resulting possible narrow CW. The scatter fraction measured with the small RF Tx/Rx coil is comparable to the lowest values obtained for other preclinical systems (Goertzen *et al* 2012). The slight increase of the determined scatter fraction towards high activity can be explained by fluctuations in the evaluation method, e.g. random estimation or by pile-up-related effects (Stearns and Manjeshwar 2011).

The NECR peak values strongly depended on the chosen trig and val. For the used raw DPC sensor data mode, the GbE interfaces were the bottleneck of the maximum rate of sensor data that the system was able to deliver. If the limit is reached, hit data is statistically thrown away on individual SDMs which are in saturation, and this explains why the system showed a stronger decline in the trues sensitivity and NECR when reaching this limit. Measurements with one or more SDMs in saturation should be avoided as hit data is lost depending on the number of hits recorded per SDM. This can lead to activity-dependent loss rates for SDMs—especially if the activity is not symmetrically distributed around the isocenter of the scanner. Artifacts in the estimated trues rate, randoms rate, scatter fraction and NECR for measurements with SDMs in saturation are therefore most likely caused by data loss. A peak NECR of ~280 kcps–320 kcps at activity of 30 MBq–55 MBq delivers good sensitivity for most preclinical applications. In addition, the processing and compression of sensor data in the firmware should allow the peak NECRs to be pushed to higher values at higher activity until the detector-stack-to-SDM-communication interface or the sensor itself are saturated.

The sensitivity of trig 1 was significantly lower compared to the higher trig for all $T_{op}$ using the COG-ACE crystal identification method. With the employed light sharing, a statistical crystal identification method, such as maximum-likelihood estimation, could be used to recover the sensitivity loss of trig 1. This, however, would probably lead to a degradation of $\Delta E/E$ and CRT performance.

### 5.5. Image spatial resolution and benefit of TOF

In the 2D slice of the mouse-sized hot-rod phantom, the scanner was able to separate the 0.8 mm rod region, with the exception of the two rods closest to the center of the phantom, which were not resolved. All 0.9 mm rods were separable and clearly identifiable, and although not yet determined, according to the NEMA NU 4 standard, the spatial resolution of the scanner seems comparable to the best values reported for other preclinical systems (Goertzen *et al* 2012). Trig 1 and trig 3 delivered almost the same peak-to-valley value, which indicates that the reconstruction was most likely not limited by statistics. Furthermore, no observable benefit of reconstruction using the TOF information for either the trig 1 or the trig 3 measurement could be shown. This is not surprising, as the CRT of trig 1 (~260 ps) translates to a spatial resolution along the LOR of 39 mm, which is still larger than the mouse-sized hot-rod phantom. The better CRT may therefore be used to reduce the random fraction at the cost of sensitivity (Eriksson and Conti 2015).

For the rabbit-sized phantom, trig 1 with TOF information showed clear benefits due to the large object diameter. In particular, in regions of overlapping horizontal and vertical gaps of the system's geometry that are parallel to the symmetry axes of the Cartesian grid of the phantom, TOF information helped to improve the image quality (figure 13). Trig 3 did not show an observable visual improvement when using TOF information for the reconstruction, but the signal-to-background evaluation indicates the benefit of using TOF information even for trig 3. The trig 3 reconstructions using the complete statistics, which were more than threefold the number of coincidences compared to trig 1, were not able to deliver the level of image quality or signal-to-background values of trig 1 reconstructions using TOF information.

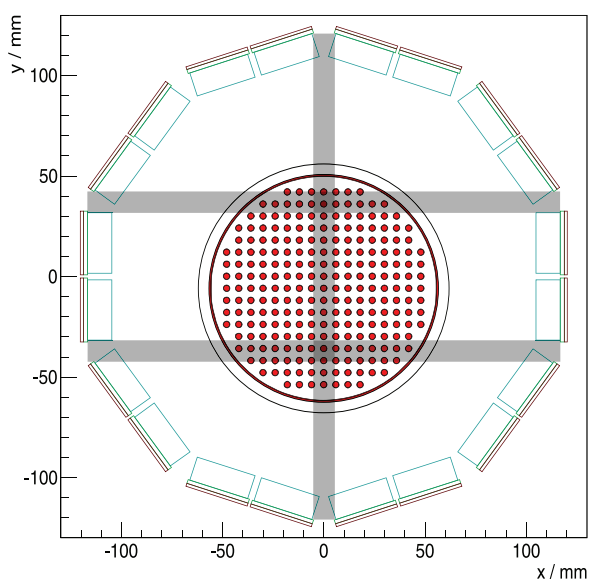

**Figure 13.** Transversal cross section through the gantry. The LYSO crystal array outlines are drawn in light blue, and the light guide in green; the sensor tile is sketched in red. The rabbit-sized phantom is shown at the measured untilted position, which is not centered along the *y*-axis. The gaps in the horizontal and vertical direction are plotted in gray, parallel to the symmetry axes of the Cartesian grid of the phantom and intersecting the activity distribution.

**Table 7.** Recommended measurement settings when using the Hyperion-II$^D$ scanner to capture the raw DPC sensor data and approximate values for the expected performance.

| Scenario | ~$T_{op}$ | $V_{ov}$ | Trig | Val | $\Delta E/E$ | CRT | ~NECR peak |
|---|---|---|---|---|---|---|---|
| Low activity (<30 MBq) | <15 °C | 2.5 V | 2/3 | 28ph | ~12.7% | 440 ps/550 ps | 280 kcps(40 MBq) |
| High activity (<60 MBq) | <15 °C | 2.5 V | 2/3 | 37ph | ~12.7% | 440 ps/550 ps | 300 kcps(55 MBq) |
| High temperature | <25 °C | 2.5 V | 4 | 28ph | ~13.0% | 1300 ps | Not measured |
| Large object ($\varnothing \gg 40$ mm) | <10 °C | 2.5 V | 1 | 28ph | ~12.8% | 260 ps | 120 kcps(25 MBq) |

Even though the rotated phantom measurement showed a more homogeneous image quality over the whole area of the phantom, there are still some artifacts visible around the central region (figure 12). They are significantly reduced by the trig 1 measurement with TOF information.

The measurements using the rabbit-sized phantom suggest an improvement in image quality due to TOF information which cannot be compensated for by a non-TOF measurement with higher statistics. The cause of this might be the non-homogeneous image spatial resolution of the scanner, which has a directional dependency. It must also be mentioned that the phantom with rods on a Cartesian grid has a very artificial distribution of activity— a smoother activity distribution containing only some lesions might show different behavior in the image quality.

A detailed evaluation of the image quality in terms of SNR is ongoing and will be part of the full NEMA NU 4 characterization.

## 6. Conclusion

The robust COG-ACE algorithm presented in Schug *et al* (2015b) was applied to the raw DPC sensor data captured with the Hyperion-II$^D$ scanner using a wide range of operating parameters. The system and applied algorithm showed very stable PET performance results under a wide range of these parameters. The presented initial evaluation of the PET performance results in a good understanding of the system and its behavior under a variety of parameters.

Aggressive voltage settings only have very minor benefits for the energy and timing performance of the system compared to the conservative choice of $V_{ov} = 2.5$ V (relative change <10%). Higher voltages can be used to increase the sensitivity for trig 4 and high val, lower trig and val, on the other hand, lose sensitivity at aggressive voltage settings. We therefore conclude that it is not beneficial to use aggressive $V_{ov} > 2.5$ V for the presented imaging applications.

Low $T_{op}$ are beneficial for all operation parameters. Thus, to apply trig 1, the DPCs should be operated at $T_{op} \lesssim 10\,°C$, as otherwise, the dead time causes a significant loss of sensitivity—especially in the presented application using light sharing. Even lower temperatures are preferable to keep the sensitivity loss at a minimum. Trig 2 and 3 still deliver CRTs of about 440 ps and 550 ps, while being more robust to higher temperatures—these should be the preferred choice for most preclinical applications. We were able to show that the CRT of ∼260 ps for trig 1 outperforms the trig 3 setting for a rabbit-sized activity distribution. On the other hand, for a mouse-sized phantom, the TOF information of trig 1 did not help to improve the image quality noticeably. We therefore conclude that trig 1 should only be the preferred choice if the diameter of the activity distribution is large (≫40 mm) in order to benefit from the TOF information.

Using light sharing, one should use low val to increase the probability that all the DPCs required for crystal identification (Schug *et al* 2015b) are validated. On the other hand, low val leads to the generation of many DPC hits and the large amount of data leads to a saturation of the GbE interfaces at activities of about 25 MBq–30 MBq. If high-activity measurements are performed (>50 MBq–60 MBq), high val give a better NECR performance as the current limit given by the GbE interfaces is shifted to a higher activity. The compression or processing of raw DPC sensor data on the detector stack and/or SDM level will help to remove this bottleneck.

For the Hyperion-II$^D$ scanner in the presented configuration, equipped with six detector stacks per SDM and using the raw DPC sensor data mode, we suggest the measurement settings listed in table 7.

## 7. Outlook

Based on the findings in this paper, a few sets of parameters will be selected and a scanner characterization will be conducted following the full NEMA NU 4 standard. In addition, the TOF benefit will be evaluated in terms of SNR gain using standardized phantoms. The interference study between the Hyperion-II$^D$ platform and a 3 T MRI is ongoing (Wehner *et al* 2014, Schug *et al* 2015a, Wehner *et al* 2015, Weissler *et al* 2015).

With the unique possibility of capturing raw DPC sensor data, the Hyperion-II$^D$ scanner will allow a wide range of different processing methods to be studied, since they can be implemented in flexible software-based frameworks, which will allow these methods and their parameters to be evaluated for exactly the same PET measurement.

## Acknowledgments

This work was supported by the European Community Seventh Framework Programme, project number 241711: SUB nanosecond leverage in PET/MR Imaging (SUBLIMA).

The project 'ForSaTum' is co-funded by the European Union (European Regional Development Fund—Investing in your future) and the German federal state North Rhine-Westphalia (NRW).

The Centre of Excellence in Medical Engineering, funded by the Wellcome Trust and EPSRC under grant number WT 088641/Z/09/Z.

The presented work is financially supported by Philips Research Europe, Aachen, Germany.

# Initial PET Performance Evaluation of a Preclinical Insert for PET/MRI with Digital SiPM Technology

**Supplemental data**

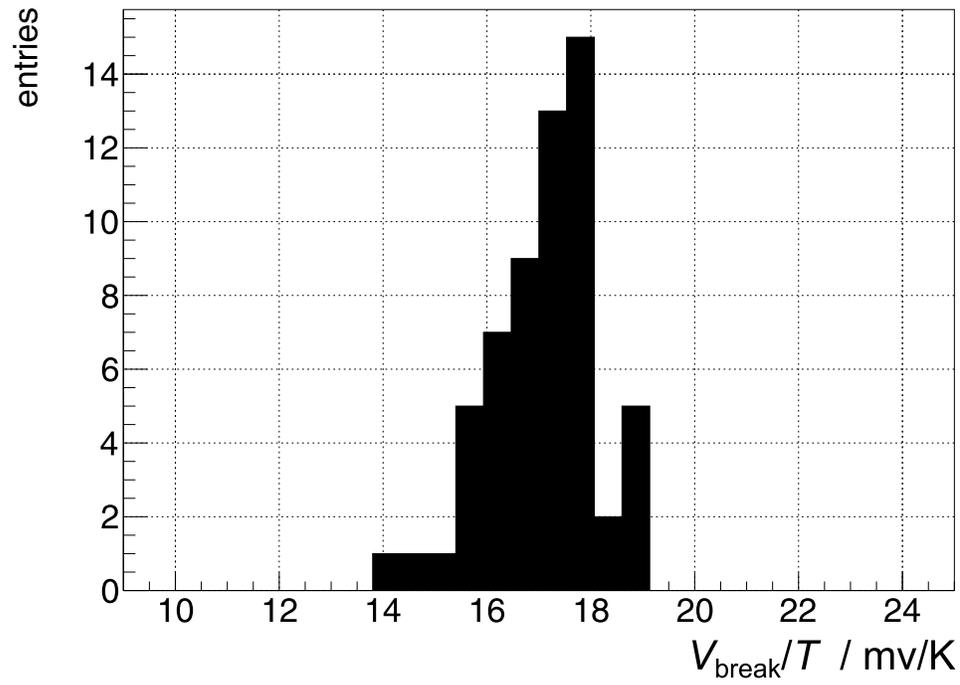

**Figure S1.** Histogram of the dependence of the breakdown voltage on temperature of all tiles of the system, except for one tile with a faulty temperature sensor.

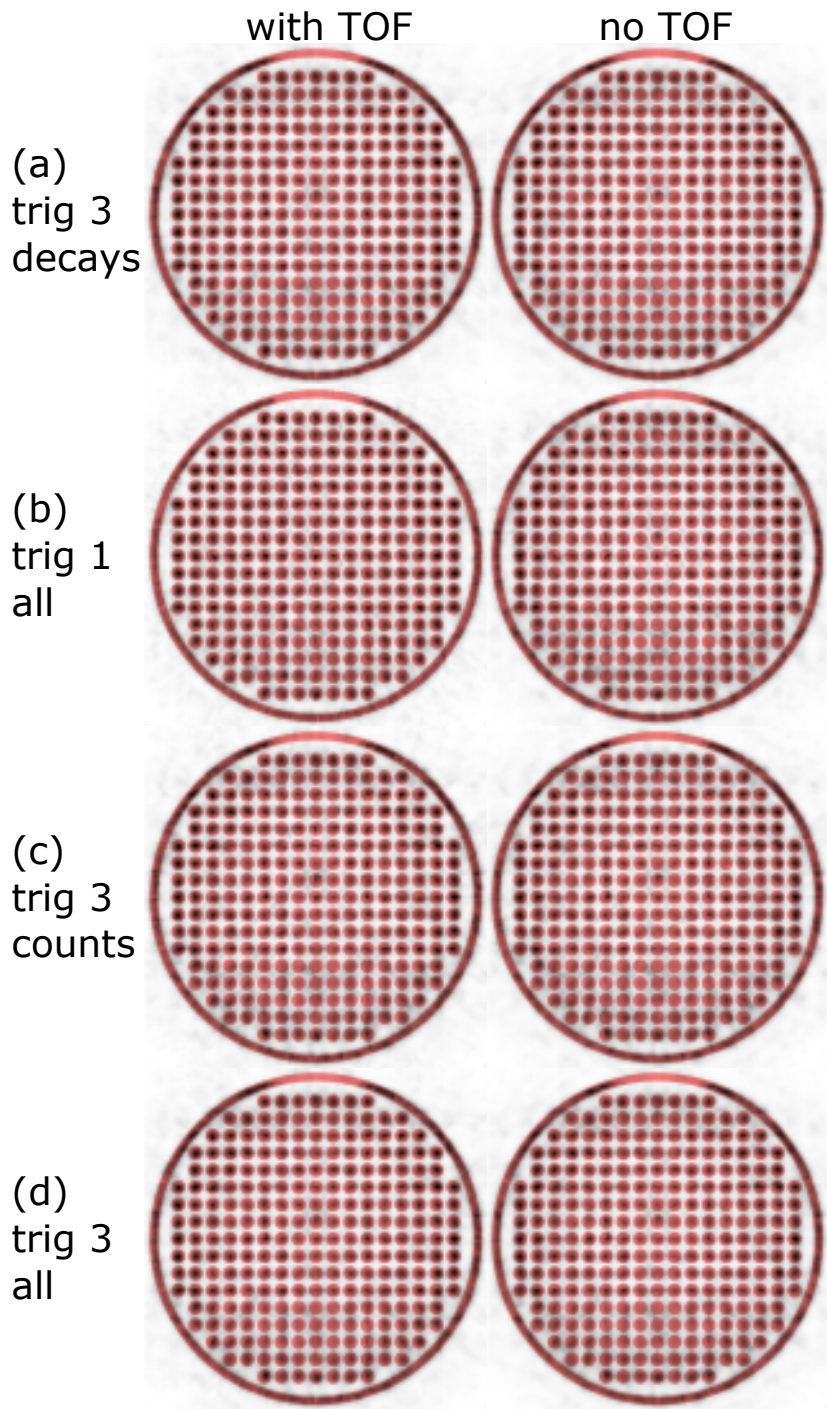

**Figure S2.** Transversal slices through the reconstructed rabbit-sized phantom as shown in ??. The signal region used for the signal-to-background evaluation is indicated in transparent red.

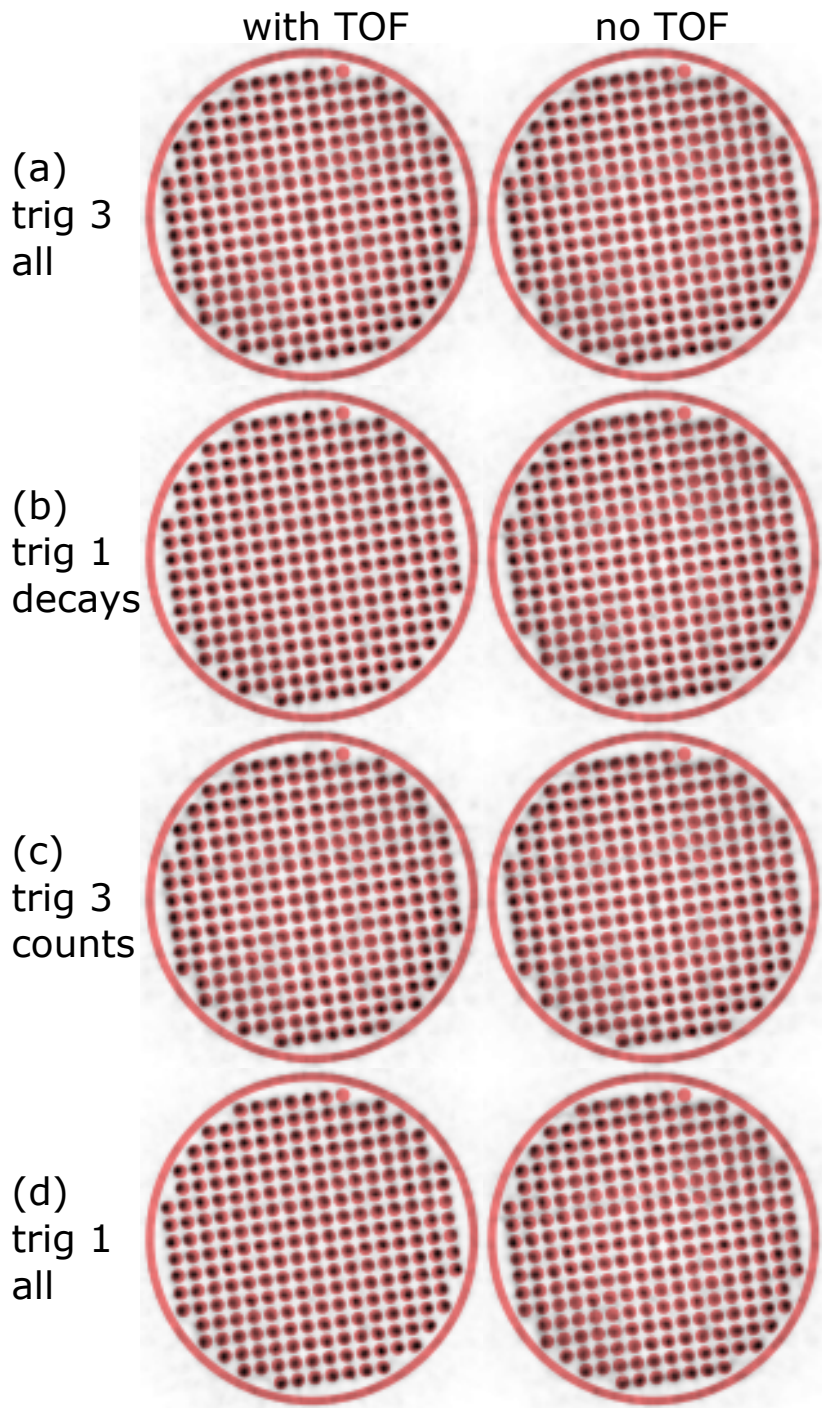

**Figure S3.** Transversal slices through the reconstructed rabbit-sized phantom measured in the tilted position as shown in **??**. The signal region used for the signal-to-background evaluation is indicated in transparent red.

Table S1: All measurements performed with the 5 $^{22}$Na point like sources (distribution and activities listed in ?? and constant $V_{\text{bias}}$. A measurement id is used to identify a specific measurement (meas.). The measurement parameters: cooling temperature ($T_{\text{C}}$), overvoltage ($V_{\text{ov}}$), trigger scheme (trig), validation scheme (val) and the used energy window (EW) are stated for each evaluation of a measurement. The energy resolution and CRT are stated as FWHM and the sensitivity is calculated from the ratio of the prompt rate and the activity of the point sources corrected for the branching ratio of the $^{22}$Na $\beta^+$ decay of 0.906

| meas. | $T_{\text{C}}$ | system $T_{\text{op}}$ | $V_{\text{ov}}$ | trig | val | EW | $\Delta E/E$ | CRT | sens. |
|---|---|---|---|---|---|---|---|---|---|
| 1 | −5 °C | 4.48 ± 1.45 °C | 2.5 V | 1 | 17ph | NE | 12.45 % | 258 ps | 0.66 % |
| 2 | −5 °C | 4.48 ± 1.45 °C | 2.5 V | 1 | 17ph | WE | 12.47 % | 282 ps | 1.43 % |
| 3 | −5 °C | 4.62 ± 1.47 °C | 2.8 V | 1 | 17ph | NE | 12.74 % | 259 ps | 0.56 % |
| 4 | −5 °C | 4.62 ± 1.47 °C | 2.8 V | 1 | 17ph | WE | 12.75 % | 286 ps | 1.27 % |
| 5 | −5 °C | 4.73 ± 1.48 °C | 2.9 V | 1 | 17ph | NE | 13.86 % | 262 ps | 0.48 % |
| 6 | −5 °C | 4.73 ± 1.48 °C | 2.9 V | 1 | 17ph | WE | 13.92 % | 295 ps | 1.21 % |
| 7 | −5 °C | 3.28 ± 1.17 °C | 2.5 V | 2 | 17ph | NE | 12.39 % | 432 ps | 1.13 % |
| 8 | −5 °C | 3.28 ± 1.17 °C | 2.5 V | 2 | 17ph | WE | 12.40 % | 481 ps | 2.38 % |
| 9 | −5 °C | 3.28 ± 1.16 °C | 2.5 V | 2 | 37ph | NE | 12.40 % | 429 ps | 1.03 % |
| 10 | −5 °C | 3.28 ± 1.16 °C | 2.5 V | 2 | 37ph | WE | 12.42 % | 472 ps | 1.99 % |
| 11 | −5 °C | 3.21 ± 1.15 °C | 2.5 V | 2 | 52ph | NE | 12.35 % | 454 ps | 0.78 % |
| 12 | −5 °C | 3.21 ± 1.15 °C | 2.5 V | 2 | 52ph | WE | 12.36 % | 489 ps | 1.36 % |
| 13 | −5 °C | 3.35 ± 1.21 °C | 2.8 V | 2 | 17ph | NE | 12.32 % | 418 ps | 1.10 % |
| 14 | −5 °C | 3.35 ± 1.21 °C | 2.8 V | 2 | 17ph | WE | 12.34 % | 467 ps | 2.32 % |
| 15 | −5 °C | 3.41 ± 1.20 °C | 2.9 V | 2 | 17ph | NE | 12.31 % | 415 ps | 1.07 % |
| 16 | −5 °C | 3.41 ± 1.20 °C | 2.9 V | 2 | 17ph | WE | 12.33 % | 463 ps | 2.27 % |
| 17 | −5 °C | 3.53 ± 1.16 °C | 3 V | 2 | 17ph | NE | 14.10 % | 416 ps | 0.93 % |
| 18 | −5 °C | 3.53 ± 1.16 °C | 3 V | 2 | 17ph | WE | 14.10 % | 467 ps | 2.20 % |
| 19 | −5 °C | 3.34 ± 1.18 °C | 2.5 V | 3 | 17ph | NE | 12.39 % | 538 ps | 1.14 % |
| 20 | −5 °C | 3.34 ± 1.18 °C | 2.5 V | 3 | 17ph | WE | 12.41 % | 598 ps | 2.38 % |
| 21 | −5 °C | 3.17 ± 1.15 °C | 2.5 V | 4 | 17ph | NE | 12.50 % | 1241 ps | 1.16 % |
| 22 | −5 °C | 3.17 ± 1.15 °C | 2.5 V | 4 | 17ph | WE | 12.55 % | 1394 ps | 2.43 % |
| 23 | 5 °C | 13.41 ± 1.32 °C | 2.5 V | 1 | 17ph | NE | 12.81 % | 272 ps | 0.54 % |
| 24 | 5 °C | 13.41 ± 1.32 °C | 2.5 V | 1 | 17ph | WE | 12.72 % | 298 ps | 0.80 % |
| 25 | 5 °C | 11.67 ± 1.05 °C | 2.5 V | 2 | 17ph | NE | 12.52 % | 440 ps | 1.10 % |
| 26 | 5 °C | 11.67 ± 1.05 °C | 2.5 V | 2 | 17ph | WE | 12.56 % | 490 ps | 2.31 % |
| 27 | 5 °C | 11.67 ± 1.06 °C | 2.5 V | 2 | 37ph | NE | 12.55 % | 437 ps | 0.97 % |
| 28 | 5 °C | 11.67 ± 1.06 °C | 2.5 V | 2 | 37ph | WE | 12.57 % | 480 ps | 1.87 % |
| 29 | 5 °C | 11.67 ± 1.06 °C | 2.5 V | 2 | 52ph | NE | 12.58 % | 436 ps | 0.73 % |
| 30 | 5 °C | 11.67 ± 1.06 °C | 2.5 V | 2 | 52ph | WE | 12.59 % | 476 ps | 1.32 % |
| 31 | 5 °C | 11.67 ± 1.06 °C | 2.5 V | 3 | 17ph | NE | 12.65 % | 552 ps | 1.20 % |
| 32 | 5 °C | 11.67 ± 1.06 °C | 2.5 V | 3 | 17ph | WE | 12.57 % | 605 ps | 2.33 % |
| 33 | 5 °C | 11.51 ± 1.03 °C | 2.5 V | 4 | 17ph | NE | 12.74 % | 1268 ps | 1.23 % |
| 34 | 5 °C | 11.51 ± 1.03 °C | 2.5 V | 4 | 17ph | WE | 12.70 % | 1399 ps | 2.39 % |
| 35 | 15 °C | 21.03 ± 1.06 °C | 2.5 V | 2 | 17ph | NE | 12.75 % | 450 ps | 1.00 % |

Table S1 – continued from previous page

| meas. | $T_{\mathrm{C}}$ | system $T_{\mathrm{op}}$ | $V_{\mathrm{ov}}$ | trig | val | EW | $\Delta E/E$ | CRT | sens. |
|---|---|---|---|---|---|---|---|---|---|
| 36 | 15 °C | 21.03 ± 1.06 °C | 2.5 V | 2 | 17ph | WE | 12.75 % | 499 ps | 2.08 % |
| 37 | 15 °C | 21.30 ± 1.10 °C | 2.5 V | 2 | 37ph | NE | 12.74 % | 451 ps | 0.84 % |
| 38 | 15 °C | 21.30 ± 1.10 °C | 2.5 V | 2 | 37ph | WE | 12.78 % | 497 ps | 1.60 % |
| 39 | 15 °C | 21.28 ± 1.10 °C | 2.5 V | 2 | 52ph | NE | 12.81 % | 450 ps | 0.61 % |
| 40 | 15 °C | 21.28 ± 1.10 °C | 2.5 V | 2 | 52ph | WE | 12.82 % | 492 ps | 1.08 % |
| 41 | 15 °C | 21.18 ± 1.09 °C | 2.8 V | 2 | 17ph | NE | 12.63 % | 442 ps | 0.92 % |
| 42 | 15 °C | 21.18 ± 1.09 °C | 2.8 V | 2 | 17ph | WE | 12.65 % | 497 ps | 1.94 % |
| 43 | 15 °C | 21.40 ± 1.13 °C | 2.8 V | 2 | 37ph | NE | 12.64 % | 436 ps | 0.86 % |
| 44 | 15 °C | 21.40 ± 1.13 °C | 2.8 V | 2 | 37ph | WE | 12.66 % | 484 ps | 1.67 % |
| 45 | 15 °C | 21.36 ± 1.11 °C | 2.8 V | 2 | 52ph | NE | 12.68 % | 434 ps | 0.67 % |
| 46 | 15 °C | 21.36 ± 1.11 °C | 2.8 V | 2 | 52ph | WE | 12.71 % | 476 ps | 1.22 % |
| 47 | 15 °C | 21.24 ± 1.10 °C | 2.9 V | 2 | 17ph | NE | 12.62 % | 440 ps | 0.90 % |
| 48 | 15 °C | 21.24 ± 1.10 °C | 2.9 V | 2 | 17ph | WE | 12.64 % | 495 ps | 1.90 % |
| 49 | 15 °C | 21.41 ± 1.13 °C | 2.9 V | 2 | 37ph | NE | 12.63 % | 435 ps | 0.86 % |
| 50 | 15 °C | 21.41 ± 1.13 °C | 2.9 V | 2 | 37ph | WE | 12.65 % | 482 ps | 1.68 % |
| 51 | 15 °C | 21.40 ± 1.12 °C | 2.9 V | 2 | 52ph | NE | 12.64 % | 430 ps | 0.69 % |
| 52 | 15 °C | 21.40 ± 1.12 °C | 2.9 V | 2 | 52ph | WE | 12.66 % | 472 ps | 1.26 % |
| 53 | 15 °C | 21.31 ± 1.12 °C | 3 V | 2 | 17ph | NE | 12.60 % | 439 ps | 0.87 % |
| 54 | 15 °C | 21.31 ± 1.12 °C | 3 V | 2 | 17ph | WE | 12.62 % | 495 ps | 1.85 % |
| 55 | 15 °C | 21.47 ± 1.14 °C | 3 V | 2 | 37ph | NE | 12.60 % | 433 ps | 0.85 % |
| 56 | 15 °C | 21.47 ± 1.14 °C | 3 V | 2 | 37ph | WE | 12.62 % | 481 ps | 1.68 % |
| 57 | 15 °C | 21.47 ± 1.13 °C | 3 V | 2 | 52ph | NE | 12.62 % | 425 ps | 0.70 % |
| 58 | 15 °C | 21.47 ± 1.13 °C | 3 V | 2 | 52ph | WE | 12.64 % | 470 ps | 1.29 % |
| 59 | 15 °C | 21.22 ± 1.07 °C | 2.5 V | 3 | 17ph | NE | 12.74 % | 562 ps | 1.02 % |
| 60 | 15 °C | 21.22 ± 1.07 °C | 2.5 V | 3 | 17ph | WE | 12.76 % | 632 ps | 2.14 % |
| 61 | 15 °C | 21.25 ± 1.07 °C | 2.8 V | 3 | 17ph | NE | 12.64 % | 544 ps | 0.99 % |
| 62 | 15 °C | 21.25 ± 1.07 °C | 2.8 V | 3 | 17ph | WE | 12.67 % | 613 ps | 2.09 % |
| 63 | 15 °C | 21.29 ± 1.08 °C | 2.9 V | 3 | 17ph | NE | 12.62 % | 540 ps | 0.97 % |
| 64 | 15 °C | 21.29 ± 1.08 °C | 2.9 V | 3 | 17ph | WE | 12.65 % | 608 ps | 2.06 % |
| 65 | 15 °C | 21.33 ± 1.09 °C | 3 V | 3 | 17ph | NE | 12.60 % | 537 ps | 0.95 % |
| 66 | 15 °C | 21.33 ± 1.09 °C | 3 V | 3 | 17ph | WE | 12.62 % | 604 ps | 2.02 % |
| 67 | 15 °C | 20.60 ± 0.95 °C | 2.5 V | 4 | 17ph | NE | 12.85 % | 1300 ps | 1.11 % |
| 68 | 15 °C | 20.60 ± 0.95 °C | 2.5 V | 4 | 17ph | WE | 12.88 % | 1469 ps | 2.34 % |
| 69 | 15 °C | 20.66 ± 0.97 °C | 2.8 V | 4 | 17ph | NE | 12.75 % | 1234 ps | 1.12 % |
| 70 | 15 °C | 20.66 ± 0.97 °C | 2.8 V | 4 | 17ph | WE | 12.78 % | 1399 ps | 2.36 % |
| 71 | 15 °C | 20.73 ± 0.99 °C | 2.9 V | 4 | 17ph | NE | 12.73 % | 1214 ps | 1.12 % |
| 72 | 15 °C | 20.73 ± 0.99 °C | 2.9 V | 4 | 17ph | WE | 12.76 % | 1375 ps | 2.37 % |
| 73 | 15 °C | 20.79 ± 0.99 °C | 3 V | 4 | 17ph | NE | 12.69 % | 1189 ps | 1.11 % |
| 74 | 15 °C | 20.79 ± 0.99 °C | 3 V | 4 | 17ph | WE | 12.72 % | 1344 ps | 2.36 % |

Table S2: All measurements performed with the 5 $^{22}$Na point like sources (distribution and activities listed in ?? and constant $V_{\text{ov}}$. A measurement id is used to identify a specific measurement (meas.). The measurement parameters: cooling temperature ($T_{\text{C}}$), system operating temperature ($T_{\text{op}}$), overvoltage ($V_{\text{ov}}$), trigger scheme (trig), validation scheme (val) and the used energy window (EW) are stated for each evaluation of a measurement. The energy resolution and CRT are stated as FWHM and the sensitivity is calculated from the ratio of the prompt rate and the activity of the point sources corrected for the branching ratio of the $^{22}$Na $\beta^+$ decay of 0.906

| meas. | $T_{\text{C}}$ | system $T_{\text{op}}$ | $V_{\text{ov}}$ | trig | val | EW | $\Delta E/E$ | CRT | sens. |
|---|---|---|---|---|---|---|---|---|---|
| 1 | $-5\,°\text{C}$ | $4.86 \pm 1.42\,°\text{C}$ | $2.5\,\text{V}$ | 1 | 17ph | NE | $12.62\,\%$ | $262\,\text{ps}$ | $0.65\,\%$ |
| 2 | $-5\,°\text{C}$ | $4.86 \pm 1.42\,°\text{C}$ | $2.5\,\text{V}$ | 1 | 17ph | WE | $12.62\,\%$ | $288\,\text{ps}$ | $1.40\,\%$ |
| 3 | $-5\,°\text{C}$ | $4.18 \pm 1.27\,°\text{C}$ | $2.5\,\text{V}$ | 3 | 17ph | NE | $12.57\,\%$ | $576\,\text{ps}$ | $1.03\,\%$ |
| 4 | $-5\,°\text{C}$ | $4.18 \pm 1.27\,°\text{C}$ | $2.5\,\text{V}$ | 3 | 17ph | WE | $12.57\,\%$ | $638\,\text{ps}$ | $2.15\,\%$ |
| 5 | $-5\,°\text{C}$ | $3.98 \pm 1.24\,°\text{C}$ | $2.5\,\text{V}$ | 4 | 17ph | NE | $12.69\,\%$ | $1336\,\text{ps}$ | $1.04\,\%$ |
| 6 | $-5\,°\text{C}$ | $3.98 \pm 1.24\,°\text{C}$ | $2.5\,\text{V}$ | 4 | 17ph | WE | $12.71\,\%$ | $1496\,\text{ps}$ | $2.18\,\%$ |
| 7 | $0\,°\text{C}$ | $9.24 \pm 1.44\,°\text{C}$ | $2.5\,\text{V}$ | 1 | 17ph | NE | $12.69\,\%$ | $267\,\text{ps}$ | $0.51\,\%$ |
| 8 | $0\,°\text{C}$ | $9.24 \pm 1.44\,°\text{C}$ | $2.5\,\text{V}$ | 1 | 17ph | WE | $12.69\,\%$ | $294\,\text{ps}$ | $1.11\,\%$ |
| 9 | $0\,°\text{C}$ | $8.00 \pm 1.12\,°\text{C}$ | $2.5\,\text{V}$ | 3 | 17ph | NE | $12.62\,\%$ | $576\,\text{ps}$ | $1.02\,\%$ |
| 10 | $0\,°\text{C}$ | $8.00 \pm 1.12\,°\text{C}$ | $2.5\,\text{V}$ | 3 | 17ph | WE | $12.62\,\%$ | $637\,\text{ps}$ | $2.14\,\%$ |
| 11 | $0\,°\text{C}$ | $7.92 \pm 1.11\,°\text{C}$ | $2.5\,\text{V}$ | 4 | 17ph | NE | $12.74\,\%$ | $1334\,\text{ps}$ | $1.04\,\%$ |
| 12 | $0\,°\text{C}$ | $7.92 \pm 1.11\,°\text{C}$ | $2.5\,\text{V}$ | 4 | 17ph | WE | $12.76\,\%$ | $1495\,\text{ps}$ | $2.17\,\%$ |
| 13 | $5\,°\text{C}$ | $13.77 \pm 1.35\,°\text{C}$ | $2.5\,\text{V}$ | 1 | 17ph | NE | $12.77\,\%$ | $307\,\text{ps}$ | $0.34\,\%$ |
| 14 | $5\,°\text{C}$ | $13.77 \pm 1.35\,°\text{C}$ | $2.5\,\text{V}$ | 1 | 17ph | WE | $12.77\,\%$ | $340\,\text{ps}$ | $0.76\,\%$ |
| 15 | $5\,°\text{C}$ | $11.82 \pm 0.97\,°\text{C}$ | $2.5\,\text{V}$ | 3 | 17ph | NE | $12.66\,\%$ | $575\,\text{ps}$ | $1.01\,\%$ |
| 16 | $5\,°\text{C}$ | $11.82 \pm 0.97\,°\text{C}$ | $2.5\,\text{V}$ | 3 | 17ph | WE | $12.67\,\%$ | $637\,\text{ps}$ | $2.12\,\%$ |
| 17 | $5\,°\text{C}$ | $11.85 \pm 0.99\,°\text{C}$ | $2.5\,\text{V}$ | 4 | 17ph | NE | $12.79\,\%$ | $1332\,\text{ps}$ | $1.04\,\%$ |
| 18 | $5\,°\text{C}$ | $11.85 \pm 0.99\,°\text{C}$ | $2.5\,\text{V}$ | 4 | 17ph | WE | $12.80\,\%$ | $1491\,\text{ps}$ | $2.17\,\%$ |
| 19 | $10\,°\text{C}$ | $16.27 \pm 0.95\,°\text{C}$ | $2.5\,\text{V}$ | 4 | 17ph | NE | $12.84\,\%$ | $1326\,\text{ps}$ | $1.03\,\%$ |
| 20 | $10\,°\text{C}$ | $16.27 \pm 0.95\,°\text{C}$ | $2.5\,\text{V}$ | 4 | 17ph | WE | $12.86\,\%$ | $1488\,\text{ps}$ | $2.16\,\%$ |
| 21 | $15\,°\text{C}$ | $20.49 \pm 0.85\,°\text{C}$ | $2.5\,\text{V}$ | 3 | 17ph | NE | $12.72\,\%$ | $563\,\text{ps}$ | $0.94\,\%$ |
| 22 | $15\,°\text{C}$ | $20.49 \pm 0.85\,°\text{C}$ | $2.5\,\text{V}$ | 3 | 17ph | WE | $12.72\,\%$ | $625\,\text{ps}$ | $2.00\,\%$ |
| 23 | $15\,°\text{C}$ | $20.56 \pm 0.88\,°\text{C}$ | $2.5\,\text{V}$ | 4 | 17ph | NE | $12.84\,\%$ | $1280\,\text{ps}$ | $1.02\,\%$ |
| 24 | $15\,°\text{C}$ | $20.56 \pm 0.88\,°\text{C}$ | $2.5\,\text{V}$ | 4 | 17ph | WE | $12.85\,\%$ | $1435\,\text{ps}$ | $2.16\,\%$ |

Table S3: All measurements performed with the mouse-sized scatter phantom. A measurement id is used to identify a specific measurement (meas.). The measurement parameters: source activity (activity), system operating temperature ($T_{op}$) for a cooling temperature of $T_C = 0\,°C$, trigger scheme (trig), validation scheme (val) and the used energy window (EW) are stated for each evaluation of a measurement. The energy resolution and CRT are stated as FWHM. The prompt rate (prompts), random fraction (randoms), scatter fraction (scatter), trues sensitivity (sens) and NECR are evaluated following the NEMA NU4 standard.

| meas. | activity | system $T_{op}$ | trig | val | EW | $\Delta E/E$ | CRT | prompts | randoms | scatter | sens | NECR |
|---|---|---|---|---|---|---|---|---|---|---|---|---|
| 1 | 102.60 MBq | 9.23 ± 1.35 °C | 3 | 28ph | NE | 13.60 % | 566 ps | 136.18 kcps | 2.92 % | 9.79 % | 0.12 % | 105.04 kcps |
| 2 | 102.60 MBq | 9.23 ± 1.35 °C | 3 | 28ph | WE | 13.64 % | 618 ps | 276.03 kcps | 3.74 % | 16.66 % | 0.22 % | 179.88 kcps |
| 3 | 96.37 MBq | 9.00 ± 1.32 °C | 3 | 52ph | NE | 13.46 % | 560 ps | 187.82 kcps | 3.58 % | 9.48 % | 0.17 % | 144.06 kcps |
| 4 | 96.37 MBq | 9.00 ± 1.32 °C | 3 | 52ph | WE | 13.48 % | 600 ps | 342.48 kcps | 4.30 % | 15.59 % | 0.29 % | 226.48 kcps |
| 5 | 92.40 MBq | 8.98 ± 1.32 °C | 3 | 37ph | NE | 13.44 % | 561 ps | 170.55 kcps | 3.15 % | 8.10 % | 0.16 % | 135.79 kcps |
| 6 | 92.40 MBq | 8.98 ± 1.32 °C | 3 | 37ph | WE | 13.50 % | 609 ps | 329.22 kcps | 3.97 % | 16.09 % | 0.29 % | 216.54 kcps |
| 7 | 88.96 MBq | 8.97 ± 1.32 °C | 3 | 28ph | NE | 13.43 % | 563 ps | 137.80 kcps | 2.78 % | 9.13 % | 0.14 % | 108.11 kcps |
| 8 | 88.96 MBq | 8.97 ± 1.32 °C | 3 | 28ph | WE | 13.46 % | 615 ps | 278.19 kcps | 3.71 % | 20.39 % | 0.24 % | 165.98 kcps |
| 9 | 83.16 MBq | 9.23 ± 1.36 °C | 2 | 28ph | NE | 13.48 % | 451 ps | 135.51 kcps | 2.29 % | 8.41 % | 0.15 % | 108.97 kcps |
| 10 | 83.16 MBq | 9.23 ± 1.36 °C | 2 | 28ph | WE | 13.50 % | 495 ps | 269.84 kcps | 2.97 % | 17.45 % | 0.26 % | 174.93 kcps |
| 11 | 77.32 MBq | 8.77 ± 1.28 °C | 3 | 52ph | NE | 13.25 % | 556 ps | 194.38 kcps | 2.84 % | 7.49 % | 0.23 % | 157.70 kcps |
| 12 | 77.32 MBq | 8.77 ± 1.28 °C | 3 | 52ph | WE | 13.27 % | 595 ps | 347.76 kcps | 3.40 % | 14.11 % | 0.37 % | 241.72 kcps |
| 13 | 75.04 MBq | 8.65 ± 1.26 °C | 3 | 37ph | NE | 13.24 % | 557 ps | 179.94 kcps | 2.70 % | 7.88 % | 0.22 % | 145.19 kcps |
| 14 | 75.04 MBq | 8.65 ± 1.26 °C | 3 | 37ph | WE | 13.26 % | 602 ps | 340.96 kcps | 3.38 % | 15.79 % | 0.37 % | 228.12 kcps |
| 15 | 73.51 MBq | 8.69 ± 1.26 °C | 2 | 28ph | NE | 13.24 % | 449 ps | 149.02 kcps | 2.15 % | 9.04 % | 0.18 % | 118.51 kcps |
| 16 | 73.51 MBq | 8.69 ± 1.26 °C | 2 | 28ph | WE | 13.27 % | 491 ps | 296.33 kcps | 2.78 % | 18.02 % | 0.32 % | 190.14 kcps |
| 17 | 70.70 MBq | 8.79 ± 1.29 °C | 2 | 37ph | NE | 13.21 % | 447 ps | 192.42 kcps | 2.16 % | 7.61 % | 0.25 % | 157.74 kcps |

| meas. | activity | system $T_\mathrm{op}$ | trig | val | EW | $\Delta E/E$ | CRT | prompts | randoms | scatter | sens | NECR |
| --- | --- | --- | --- | --- | --- | --- | --- | --- | --- | --- | --- | --- |
| 18 | 70.70 MBq | 8.79 ± 1.29 °C | 2 | 37ph | WE | 13.24 % | 485 ps | 363.07 kcps | 2.68 % | 15.75 % | 0.42 % | 246.12 kcps |
| 19 | 67.99 MBq | 8.69 ± 1.27 °C | 3 | 28ph | NE | 13.18 % | 557 ps | 157.13 kcps | 2.37 % | 7.82 % | 0.21 % | 127.71 kcps |
| 20 | 67.99 MBq | 8.69 ± 1.27 °C | 3 | 28ph | WE | 13.21 % | 608 ps | 308.95 kcps | 3.07 % | 16.90 % | 0.37 % | 202.56 kcps |
| 21 | 66.38 MBq | 8.60 ± 1.26 °C | 3 | 52ph | NE | 13.08 % | 551 ps | 216.94 kcps | 2.42 % | 7.78 % | 0.29 % | 176.36 kcps |
| 22 | 66.38 MBq | 8.60 ± 1.26 °C | 3 | 52ph | WE | 13.09 % | 591 ps | 382.48 kcps | 2.86 % | 13.71 % | 0.48 % | 270.85 kcps |
| 23 | 64.23 MBq | 8.56 ± 1.25 °C | 3 | 37ph | NE | 13.10 % | 555 ps | 204.15 kcps | 2.33 % | 7.51 % | 0.29 % | 167.19 kcps |
| 24 | 64.23 MBq | 8.56 ± 1.25 °C | 3 | 37ph | WE | 13.13 % | 599 ps | 382.91 kcps | 2.87 % | 15.09 % | 0.49 % | 262.72 kcps |
| 25 | 62.17 MBq | 8.62 ± 1.25 °C | 2 | 28ph | NE | 13.11 % | 447 ps | 171.67 kcps | 1.88 % | 8.42 % | 0.25 % | 139.03 kcps |
| 26 | 62.17 MBq | 8.62 ± 1.25 °C | 2 | 28ph | WE | 13.14 % | 488 ps | 336.41 kcps | 2.39 % | 16.73 % | 0.44 % | 224.05 kcps |
| 27 | 59.08 MBq | 8.55 ± 1.24 °C | 2 | 37ph | NE | 13.04 % | 444 ps | 221.02 kcps | 1.78 % | 6.98 % | 0.34 % | 184.96 kcps |
| 28 | 59.08 MBq | 8.55 ± 1.24 °C | 2 | 37ph | WE | 13.06 % | 481 ps | 412.23 kcps | 2.20 % | 15.18 % | 0.58 % | 285.56 kcps |
| 29 | 56.76 MBq | 8.59 ± 1.25 °C | 2 | 52ph | NE | 12.98 % | 440 ps | 229.72 kcps | 1.67 % | 6.18 % | 0.37 % | 195.92 kcps |
| 30 | 56.76 MBq | 8.59 ± 1.25 °C | 2 | 52ph | WE | 13.00 % | 473 ps | 401.07 kcps | 2.00 % | 14.13 % | 0.60 % | 285.64 kcps |
| 31 | 55.02 MBq | 8.48 ± 1.23 °C | 3 | 28ph | NE | 13.02 % | 554 ps | 185.37 kcps | 1.98 % | 7.67 % | 0.31 % | 152.28 kcps |
| 32 | 55.02 MBq | 8.48 ± 1.23 °C | 3 | 28ph | WE | 13.05 % | 602 ps | 359.67 kcps | 2.52 % | 16.38 % | 0.54 % | 240.95 kcps |
| 33 | 52.85 MBq | 8.34 ± 1.21 °C | 3 | 37ph | NE | 12.95 % | 551 ps | 235.52 kcps | 1.86 % | 6.43 % | 0.41 % | 199.06 kcps |
| 34 | 52.85 MBq | 8.34 ± 1.21 °C | 3 | 37ph | WE | 12.97 % | 594 ps | 435.97 kcps | 2.30 % | 14.90 % | 0.69 % | 303.46 kcps |
| 35 | 51.24 MBq | 8.29 ± 1.20 °C | 3 | 52ph | NE | 12.89 % | 545 ps | 233.09 kcps | 1.75 % | 6.21 % | 0.42 % | 198.34 kcps |
| 36 | 51.24 MBq | 8.29 ± 1.20 °C | 3 | 52ph | WE | 12.90 % | 582 ps | 406.00 kcps | 2.11 % | 14.27 % | 0.67 % | 287.71 kcps |
| 37 | 48.90 MBq | 8.42 ± 1.22 °C | 2 | 28ph | NE | 12.96 % | 443 ps | 205.72 kcps | 1.47 % | 6.89 % | 0.39 % | 173.49 kcps |
| 38 | 48.90 MBq | 8.42 ± 1.22 °C | 2 | 28ph | WE | 12.98 % | 484 ps | 396.89 kcps | 1.87 % | 16.80 % | 0.66 % | 266.24 kcps |
| 39 | 47.72 MBq | 8.37 ± 1.21 °C | 2 | 37ph | NE | 12.87 % | 441 ps | 247.20 kcps | 1.40 % | 6.66 % | 0.48 % | 209.80 kcps |
| 40 | 47.72 MBq | 8.37 ± 1.21 °C | 2 | 37ph | WE | 12.91 % | 477 ps | 454.80 kcps | 1.71 % | 15.10 % | 0.80 % | 318.34 kcps |
| 41 | 46.13 MBq | 8.34 ± 1.21 °C | 2 | 52ph | NE | 12.82 % | 438 ps | 222.72 kcps | 1.32 % | 6.30 % | 0.45 % | 190.73 kcps |
| 42 | 46.13 MBq | 8.34 ± 1.21 °C | 2 | 52ph | WE | 12.84 % | 471 ps | 385.62 kcps | 1.56 % | 13.40 % | 0.71 % | 281.45 kcps |
| 43 | 43.91 MBq | 8.38 ± 1.21 °C | 2 | 17ph | NE | 12.90 % | 443 ps | 146.90 kcps | 1.32 % | 7.19 % | 0.31 % | 123.44 kcps |
| 44 | 43.91 MBq | 8.38 ± 1.21 °C | 2 | 17ph | WE | 12.92 % | 486 ps | 291.54 kcps | 1.73 % | 17.31 % | 0.54 % | 193.66 kcps |

| meas. | activity | system $T_{op}$ | trig | val | EW | $\Delta E/E$ | CRT | prompts | randoms | scatter | sens | NECR |
|---|---|---|---|---|---|---|---|---|---|---|---|---|
| 45 | 41.28 MBq | $8.29 \pm 1.20$ °C | 3 | 28ph | NE | 12.82 % | 551 ps | 226.03 kcps | 1.44 % | 6.78 % | 0.50 % | 191.21 kcps |
| 46 | 41.28 MBq | $8.29 \pm 1.20$ °C | 3 | 28ph | WE | 12.84 % | 596 ps | 429.92 kcps | 1.81 % | 15.73 % | 0.86 % | 296.06 kcps |
| 47 | 39.46 MBq | $8.23 \pm 1.18$ °C | 3 | 37ph | NE | 12.76 % | 545 ps | 250.26 kcps | 1.32 % | 6.33 % | 0.59 % | 214.18 kcps |
| 48 | 39.46 MBq | $8.23 \pm 1.18$ °C | 3 | 37ph | WE | 12.78 % | 589 ps | 456.30 kcps | 1.61 % | 14.84 % | 0.97 % | 321.89 kcps |
| 49 | 38.37 MBq | $8.17 \pm 1.17$ °C | 3 | 52ph | NE | 12.73 % | 542 ps | 197.36 kcps | 1.29 % | 6.26 % | 0.48 % | 169.24 kcps |
| 50 | 38.37 MBq | $8.17 \pm 1.17$ °C | 3 | 52ph | WE | 12.75 % | 579 ps | 339.67 kcps | 1.53 % | 13.20 % | 0.76 % | 249.16 kcps |
| 51 | 36.74 MBq | $9.37 \pm 1.36$ °C | 1 | 17ph | NE | 13.00 % | 274 ps | 83.50 kcps | 1.52 % | 6.38 % | 0.21 % | 71.12 kcps |
| 52 | 36.74 MBq | $9.37 \pm 1.36$ °C | 1 | 17ph | WE | 13.02 % | 295 ps | 167.90 kcps | 1.95 % | 17.29 % | 0.37 % | 111.18 kcps |
| 53 | 34.77 MBq | $8.52 \pm 1.28$ °C | 2 | 28ph | NE | 12.76 % | 440 ps | 236.51 kcps | 0.98 % | 6.44 % | 0.63 % | 203.22 kcps |
| 54 | 34.77 MBq | $8.52 \pm 1.28$ °C | 2 | 28ph | WE | 12.78 % | 479 ps | 445.86 kcps | 1.23 % | 16.12 % | 1.06 % | 307.22 kcps |
| 55 | 33.51 MBq | $8.26 \pm 1.21$ °C | 2 | 37ph | NE | 12.69 % | 437 ps | 225.77 kcps | 0.95 % | 7.63 % | 0.62 % | 189.28 kcps |
| 56 | 33.51 MBq | $8.26 \pm 1.21$ °C | 2 | 37ph | WE | 12.71 % | 474 ps | 409.27 kcps | 1.14 % | 14.96 % | 1.03 % | 290.29 kcps |
| 57 | 32.46 MBq | $8.16 \pm 1.19$ °C | 2 | 52ph | NE | 12.69 % | 436 ps | 172.74 kcps | 0.91 % | 6.09 % | 0.50 % | 149.74 kcps |
| 58 | 32.46 MBq | $8.16 \pm 1.19$ °C | 2 | 52ph | WE | 12.70 % | 467 ps | 295.73 kcps | 1.08 % | 13.40 % | 0.78 % | 217.63 kcps |
| 59 | 31.41 MBq | $8.16 \pm 1.17$ °C | 2 | 17ph | NE | 12.71 % | 440 ps | 184.16 kcps | 0.92 % | 6.58 % | 0.54 % | 157.97 kcps |
| 60 | 31.41 MBq | $8.16 \pm 1.17$ °C | 2 | 17ph | WE | 12.73 % | 482 ps | 358.45 kcps | 1.19 % | 17.96 % | 0.93 % | 236.57 kcps |
| 61 | 29.79 MBq | $8.09 \pm 1.16$ °C | 3 | 28ph | NE | 12.66 % | 544 ps | 222.62 kcps | 0.98 % | 6.49 % | 0.69 % | 191.12 kcps |
| 62 | 29.79 MBq | $8.09 \pm 1.16$ °C | 3 | 28ph | WE | 12.67 % | 591 ps | 418.93 kcps | 1.22 % | 15.40 % | 1.18 % | 293.64 kcps |
| 63 | 29.06 MBq | $8.03 \pm 1.15$ °C | 3 | 37ph | NE | 12.62 % | 543 ps | 203.13 kcps | 0.96 % | 6.46 % | 0.65 % | 174.56 kcps |
| 64 | 29.06 MBq | $8.03 \pm 1.15$ °C | 3 | 37ph | WE | 12.64 % | 584 ps | 367.40 kcps | 1.16 % | 14.57 % | 1.07 % | 262.84 kcps |
| 65 | 28.24 MBq | $8.01 \pm 1.14$ °C | 3 | 52ph | NE | 12.61 % | 540 ps | 155.82 kcps | 0.94 % | 6.03 % | 0.51 % | 135.18 kcps |
| 66 | 28.24 MBq | $8.01 \pm 1.14$ °C | 3 | 52ph | WE | 12.63 % | 577 ps | 265.96 kcps | 1.13 % | 13.85 % | 0.80 % | 193.59 kcps |
| 67 | 27.13 MBq | $9.12 \pm 1.32$ °C | 1 | 17ph | NE | 12.84 % | 271 ps | 89.82 kcps | 1.06 % | 6.15 % | 0.31 % | 77.54 kcps |
| 68 | 27.13 MBq | $9.12 \pm 1.32$ °C | 1 | 17ph | WE | 12.85 % | 292 ps | 178.30 kcps | 1.36 % | 16.86 % | 0.54 % | 120.46 kcps |
| 69 | 26.16 MBq | $9.44 \pm 1.42$ °C | 1 | 28ph | NE | 12.83 % | 271 ps | 96.26 kcps | 0.89 % | 6.01 % | 0.34 % | 83.61 kcps |
| 70 | 26.16 MBq | $9.44 \pm 1.42$ °C | 1 | 28ph | WE | 12.84 % | 290 ps | 183.02 kcps | 1.12 % | 15.80 % | 0.58 % | 127.31 kcps |
| 71 | 23.63 MBq | $8.05 \pm 1.16$ °C | 2 | 28ph | NE | 12.58 % | 438 ps | 185.62 kcps | 0.66 % | 8.01 % | 0.72 % | 155.15 kcps |

| meas. | activity | system $T_{\text{op}}$ | trig | val | EW | $\Delta E/E$ | CRT | prompts | randoms | scatter | sens | NECR |
|---|---|---|---|---|---|---|---|---|---|---|---|---|
| 72 | 23.63 MBq | 8.05 ± 1.16 °C | 2 | 28ph | WE | 12.59 % | 477 ps | 347.87 kcps | 0.81 % | 15.38 % | 1.24 % | 245.69 kcps |
| 73 | 22.16 MBq | 8.03 ± 1.15 °C | 2 | 37ph | NE | 12.55 % | 437 ps | 162.77 kcps | 0.61 % | 6.30 % | 0.68 % | 141.28 kcps |
| 74 | 22.16 MBq | 8.03 ± 1.15 °C | 2 | 37ph | WE | 12.59 % | 472 ps | 292.57 kcps | 0.74 % | 14.67 % | 1.12 % | 210.32 kcps |
| 75 | 21.36 MBq | 8.03 ± 1.15 °C | 2 | 17ph | NE | 12.57 % | 437 ps | 171.76 kcps | 0.59 % | 6.38 % | 0.75 % | 148.87 kcps |
| 76 | 21.36 MBq | 8.03 ± 1.15 °C | 2 | 17ph | WE | 12.59 % | 479 ps | 330.28 kcps | 0.75 % | 16.12 % | 1.29 % | 229.42 kcps |
| 77 | 20.60 MBq | 7.95 ± 1.14 °C | 3 | 28ph | NE | 12.56 % | 542 ps | 166.22 kcps | 0.67 % | 6.52 % | 0.75 % | 143.42 kcps |
| 78 | 20.60 MBq | 7.95 ± 1.14 °C | 3 | 28ph | WE | 12.57 % | 588 ps | 310.16 kcps | 0.84 % | 15.30 % | 1.27 % | 219.35 kcps |
| 79 | 19.55 MBq | 7.95 ± 1.14 °C | 3 | 37ph | NE | 12.52 % | 542 ps | 146.88 kcps | 0.64 % | 6.22 % | 0.70 % | 127.63 kcps |
| 80 | 19.55 MBq | 7.95 ± 1.14 °C | 3 | 37ph | WE | 12.53 % | 582 ps | 263.01 kcps | 0.79 % | 14.86 % | 1.14 % | 188.10 kcps |
| 81 | 19.18 MBq | 7.87 ± 1.12 °C | 3 | 52ph | NE | 12.52 % | 539 ps | 112.74 kcps | 0.63 % | 5.88 % | 0.55 % | 98.68 kcps |
| 82 | 19.18 MBq | 7.87 ± 1.12 °C | 3 | 52ph | WE | 12.53 % | 575 ps | 190.69 kcps | 0.76 % | 13.10 % | 0.86 % | 142.09 kcps |
| 83 | 18.05 MBq | 9.26 ± 1.40 °C | 1 | 17ph | NE | 12.72 % | 270 ps | 66.78 kcps | 0.70 % | 6.15 % | 0.34 % | 58.05 kcps |
| 84 | 18.05 MBq | 9.26 ± 1.40 °C | 1 | 17ph | WE | 12.75 % | 291 ps | 131.03 kcps | 0.90 % | 16.58 % | 0.60 % | 89.81 kcps |
| 85 | 17.52 MBq | 9.32 ± 1.45 °C | 1 | 28ph | NE | 12.71 % | 269 ps | 70.73 kcps | 0.59 % | 6.00 % | 0.38 % | 61.80 kcps |
| 86 | 17.52 MBq | 9.32 ± 1.45 °C | 1 | 28ph | WE | 12.73 % | 288 ps | 133.32 kcps | 0.74 % | 15.97 % | 0.64 % | 92.96 kcps |
| 87 | 13.11 MBq | 7.80 ± 1.11 °C | 3 | 28ph | NE | 12.44 % | 540 ps | 112.35 kcps | 0.43 % | 6.33 % | 0.80 % | 97.79 kcps |
| 88 | 13.11 MBq | 7.80 ± 1.11 °C | 3 | 28ph | WE | 12.48 % | 586 ps | 208.40 kcps | 0.54 % | 15.46 % | 1.34 % | 147.60 kcps |
| 89 | 12.76 MBq | 7.79 ± 1.10 °C | 3 | 37ph | NE | 12.45 % | 539 ps | 100.80 kcps | 0.42 % | 6.19 % | 0.74 % | 88.02 kcps |
| 90 | 12.76 MBq | 7.79 ± 1.10 °C | 3 | 37ph | WE | 12.47 % | 581 ps | 179.70 kcps | 0.52 % | 14.83 % | 1.19 % | 129.19 kcps |
| 91 | 12.42 MBq | 7.75 ± 1.09 °C | 3 | 52ph | NE | 12.45 % | 537 ps | 76.50 kcps | 0.41 % | 5.95 % | 0.58 % | 67.14 kcps |
| 92 | 12.42 MBq | 7.75 ± 1.09 °C | 3 | 52ph | WE | 12.46 % | 574 ps | 128.81 kcps | 0.50 % | 13.35 % | 0.90 % | 95.89 kcps |
| 93 | 11.90 MBq | 7.80 ± 1.10 °C | 2 | 17ph | NE | 12.44 % | 436 ps | 104.38 kcps | 0.33 % | 6.32 % | 0.82 % | 91.03 kcps |
| 94 | 11.90 MBq | 7.80 ± 1.10 °C | 2 | 17ph | WE | 12.45 % | 477 ps | 199.32 kcps | 0.42 % | 16.07 % | 1.40 % | 139.41 kcps |
| 95 | 11.22 MBq | 7.81 ± 1.11 °C | 2 | 28ph | NE | 12.43 % | 435 ps | 97.47 kcps | 0.31 % | 6.25 % | 0.81 % | 85.17 kcps |
| 96 | 11.22 MBq | 7.81 ± 1.11 °C | 2 | 28ph | WE | 12.45 % | 473 ps | 180.91 kcps | 0.39 % | 15.33 % | 1.36 % | 128.83 kcps |
| 97 | 10.71 MBq | 8.71 ± 1.28 °C | 1 | 17ph | NE | 12.57 % | 268 ps | 44.42 kcps | 0.40 % | 6.12 % | 0.39 % | 38.86 kcps |
| 98 | 10.71 MBq | 8.71 ± 1.28 °C | 1 | 17ph | WE | 12.56 % | 289 ps | 86.84 kcps | 0.53 % | 17.11 % | 0.67 % | 59.15 kcps |

| meas. | activity | system $T_{op}$ | trig | val | EW | $\Delta E/E$ | CRT | prompts | randoms | scatter | sens | NECR |
|---|---|---|---|---|---|---|---|---|---|---|---|---|
| 99  | 10.48 MBq | 8.91 ± 1.35 °C | 1 | 28ph | NE | 12.55 % | 268 ps | 46.37 kcps  | 0.35 % | 6.19 %  | 0.41 % | 40.54 kcps  |
| 100 | 10.48 MBq | 8.91 ± 1.35 °C | 1 | 28ph | WE | 12.56 % | 287 ps | 86.98 kcps  | 0.44 % | 15.58 % | 0.70 % | 61.52 kcps  |
| 101 | 9.17 MBq  | 7.83 ± 1.12 °C | 3 | 28ph | NE | 12.41 % | 540 ps | 80.93 kcps  | 0.30 % | 6.18 %  | 0.83 % | 70.84 kcps  |
| 102 | 9.17 MBq  | 7.83 ± 1.12 °C | 3 | 28ph | WE | 12.42 % | 586 ps | 149.43 kcps | 0.39 % | 15.58 % | 1.37 % | 105.80 kcps |
| 103 | 8.82 MBq  | 7.75 ± 1.11 °C | 3 | 37ph | NE | 12.40 % | 539 ps | 71.63 kcps  | 0.29 % | 6.08 %  | 0.76 % | 62.84 kcps  |
| 104 | 8.82 MBq  | 7.75 ± 1.11 °C | 3 | 37ph | WE | 12.42 % | 581 ps | 127.11 kcps | 0.37 % | 14.75 % | 1.23 % | 91.80 kcps  |
| 105 | 8.59 MBq  | 7.70 ± 1.10 °C | 3 | 52ph | NE | 12.39 % | 537 ps | 54.36 kcps  | 0.29 % | 5.87 %  | 0.59 % | 47.91 kcps  |
| 106 | 8.59 MBq  | 7.70 ± 1.10 °C | 3 | 52ph | WE | 12.40 % | 574 ps | 91.15 kcps  | 0.36 % | 14.05 % | 0.91 % | 66.93 kcps  |
| 107 | 8.32 MBq  | 7.74 ± 1.10 °C | 2 | 17ph | NE | 12.39 % | 435 ps | 74.71 kcps  | 0.23 % | 6.20 %  | 0.84 % | 65.44 kcps  |
| 108 | 8.32 MBq  | 7.74 ± 1.10 °C | 2 | 17ph | WE | 12.41 % | 476 ps | 142.21 kcps | 0.30 % | 16.31 % | 1.43 % | 99.09 kcps  |
| 109 | 7.87 MBq  | 7.74 ± 1.10 °C | 2 | 28ph | NE | 12.35 % | 435 ps | 70.10 kcps  | 0.22 % | 6.28 %  | 0.83 % | 61.31 kcps  |
| 110 | 7.87 MBq  | 7.74 ± 1.10 °C | 2 | 28ph | WE | 12.37 % | 473 ps | 129.66 kcps | 0.28 % | 15.57 % | 1.39 % | 91.99 kcps  |
| 111 | 6.53 MBq  | 9.86 ± 1.39 °C | 1 | 17ph | NE | 12.61 % | 267 ps | 26.13 kcps  | 0.27 % | 6.02 %  | 0.37 % | 22.96 kcps  |
| 112 | 6.53 MBq  | 9.86 ± 1.39 °C | 1 | 17ph | WE | 12.63 % | 289 ps | 50.49 kcps  | 0.35 % | 16.94 % | 0.64 % | 34.63 kcps  |
| 113 | 6.38 MBq  | 9.39 ± 1.47 °C | 1 | 28ph | NE | 12.58 % | 268 ps | 28.56 kcps  | 0.22 % | 5.68 %  | 0.42 % | 25.30 kcps  |
| 114 | 6.38 MBq  | 9.39 ± 1.47 °C | 1 | 28ph | WE | 12.60 % | 286 ps | 53.13 kcps  | 0.29 % | 15.54 % | 0.70 % | 37.72 kcps  |
| 115 | 6.17 MBq  | 8.11 ± 1.26 °C | 3 | 28ph | NE | 12.39 % | 540 ps | 56.04 kcps  | 0.21 % | 6.05 %  | 0.85 % | 49.27 kcps  |
| 116 | 6.17 MBq  | 8.11 ± 1.26 °C | 3 | 28ph | WE | 12.40 % | 585 ps | 102.65 kcps | 0.27 % | 15.61 % | 1.40 % | 72.78 kcps  |
| 117 | 6.03 MBq  | 7.90 ± 1.18 °C | 3 | 37ph | NE | 12.36 % | 539 ps | 50.01 kcps  | 0.20 % | 5.93 %  | 0.78 % | 44.09 kcps  |
| 118 | 6.03 MBq  | 7.90 ± 1.18 °C | 3 | 37ph | WE | 12.38 % | 580 ps | 88.34 kcps  | 0.26 % | 15.80 % | 1.23 % | 62.34 kcps  |
| 119 | 5.86 MBq  | 7.79 ± 1.14 °C | 3 | 52ph | NE | 12.36 % | 536 ps | 37.75 kcps  | 0.20 % | 5.83 %  | 0.61 % | 33.35 kcps  |
| 120 | 5.86 MBq  | 7.79 ± 1.14 °C | 3 | 52ph | WE | 12.38 % | 572 ps | 62.93 kcps  | 0.25 % | 13.10 % | 0.93 % | 47.31 kcps  |
| 121 | 5.63 MBq  | 7.77 ± 1.11 °C | 2 | 17ph | NE | 12.37 % | 436 ps | 52.15 kcps  | 0.16 % | 6.25 %  | 0.87 % | 45.70 kcps  |
| 122 | 5.63 MBq  | 7.77 ± 1.11 °C | 2 | 17ph | WE | 12.38 % | 476 ps | 98.84 kcps  | 0.21 % | 15.91 % | 1.47 % | 69.64 kcps  |
| 123 | 5.49 MBq  | 7.74 ± 1.10 °C | 2 | 28ph | NE | 12.34 % | 435 ps | 49.72 kcps  | 0.16 % | 6.27 %  | 0.85 % | 43.55 kcps  |
| 124 | 5.49 MBq  | 7.74 ± 1.10 °C | 2 | 28ph | WE | 12.35 % | 473 ps | 91.81 kcps  | 0.20 % | 15.33 % | 1.41 % | 65.59 kcps  |
| 125 | 4.81 MBq  | 9.07 ± 1.41 °C | 1 | 17ph | NE | 12.52 % | 268 ps | 20.71 kcps  | 0.19 % | 5.91 %  | 0.40 % | 18.27 kcps  |

| meas. | activity | system $T_\text{op}$ | trig | val | EW | $\Delta E/E$ | CRT | prompts | randoms | scatter | sens | NECR |
|---|---|---|---|---|---|---|---|---|---|---|---|---|
| 126 | 4.81 MBq | $9.07 \pm 1.41\,°\text{C}$ | 1 | 17ph | WE | 12.53 % | 288 ps | 40.02 kcps | 0.26 % | 16.44 % | 0.69 % | 27.82 kcps |
| 127 | 4.65 MBq | $9.11 \pm 1.44\,°\text{C}$ | 1 | 28ph | NE | 12.52 % | 268 ps | 21.56 kcps | 0.16 % | 5.84 % | 0.44 % | 19.06 kcps |
| 128 | 4.65 MBq | $9.11 \pm 1.44\,°\text{C}$ | 1 | 28ph | WE | 12.52 % | 286 ps | 40.07 kcps | 0.22 % | 15.76 % | 0.72 % | 28.33 kcps |
| 129 | 3.63 MBq | $7.22 \pm 1.00\,°\text{C}$ | 3 | 28ph | NE | 12.37 % | 539 ps | 33.13 kcps | 0.13 % | 6.23 % | 0.85 % | 29.06 kcps |
| 130 | 3.63 MBq | $7.22 \pm 1.00\,°\text{C}$ | 3 | 28ph | WE | 12.39 % | 584 ps | 61.73 kcps | 0.17 % | 15.34 % | 1.44 % | 44.12 kcps |
| 131 | 3.55 MBq | $7.46 \pm 1.03\,°\text{C}$ | 3 | 37ph | NE | 12.39 % | 538 ps | 29.71 kcps | 0.13 % | 6.09 % | 0.79 % | 26.14 kcps |
| 132 | 3.55 MBq | $7.46 \pm 1.03\,°\text{C}$ | 3 | 37ph | WE | 12.40 % | 580 ps | 53.02 kcps | 0.17 % | 14.93 % | 1.27 % | 38.26 kcps |
| 133 | 3.34 MBq | $7.65 \pm 1.08\,°\text{C}$ | 3 | 52ph | NE | 12.39 % | 536 ps | 21.78 kcps | 0.12 % | 5.95 % | 0.61 % | 19.23 kcps |
| 134 | 3.34 MBq | $7.65 \pm 1.08\,°\text{C}$ | 3 | 52ph | WE | 12.41 % | 573 ps | 36.51 kcps | 0.16 % | 14.28 % | 0.94 % | 26.75 kcps |
| 135 | 3.21 MBq | $7.68 \pm 1.09\,°\text{C}$ | 2 | 17ph | NE | 12.40 % | 435 ps | 29.89 kcps | 0.10 % | 6.10 % | 0.87 % | 26.31 kcps |
| 136 | 3.21 MBq | $7.68 \pm 1.09\,°\text{C}$ | 2 | 17ph | WE | 12.41 % | 477 ps | 56.77 kcps | 0.13 % | 15.81 % | 1.49 % | 40.14 kcps |
| 137 | 3.03 MBq | $7.67 \pm 1.09\,°\text{C}$ | 2 | 28ph | NE | 12.40 % | 433 ps | 27.99 kcps | 0.09 % | 6.10 % | 0.87 % | 24.63 kcps |
| 138 | 3.03 MBq | $7.67 \pm 1.09\,°\text{C}$ | 2 | 28ph | WE | 12.41 % | 473 ps | 51.70 kcps | 0.13 % | 15.48 % | 1.44 % | 36.85 kcps |
| 139 | 2.78 MBq | $9.05 \pm 1.43\,°\text{C}$ | 1 | 17ph | NE | 12.56 % | 267 ps | 12.28 kcps | 0.12 % | 6.13 % | 0.41 % | 10.80 kcps |
| 140 | 2.78 MBq | $9.05 \pm 1.43\,°\text{C}$ | 1 | 17ph | WE | 12.57 % | 288 ps | 23.70 kcps | 0.17 % | 16.76 % | 0.71 % | 16.38 kcps |
| 141 | 2.63 MBq | $9.13 \pm 1.47\,°\text{C}$ | 1 | 28ph | NE | 12.54 % | 267 ps | 12.36 kcps | 0.10 % | 5.68 % | 0.44 % | 10.98 kcps |
| 142 | 2.63 MBq | $9.13 \pm 1.47\,°\text{C}$ | 1 | 28ph | WE | 12.55 % | 286 ps | 22.98 kcps | 0.14 % | 15.43 % | 0.74 % | 16.40 kcps |
| 143 | 2.43 MBq | $7.92 \pm 1.20\,°\text{C}$ | 3 | 28ph | NE | 12.40 % | 539 ps | 22.59 kcps | 0.09 % | 6.03 % | 0.87 % | 19.91 kcps |
| 144 | 2.43 MBq | $7.92 \pm 1.20\,°\text{C}$ | 3 | 28ph | WE | 12.42 % | 587 ps | 41.50 kcps | 0.13 % | 15.71 % | 1.44 % | 29.42 kcps |
| 145 | 2.35 MBq | $7.74 \pm 1.14\,°\text{C}$ | 3 | 37ph | NE | 12.39 % | 538 ps | 19.94 kcps | 0.09 % | 6.02 % | 0.80 % | 17.58 kcps |
| 146 | 2.35 MBq | $7.74 \pm 1.14\,°\text{C}$ | 3 | 37ph | WE | 12.41 % | 581 ps | 35.25 kcps | 0.12 % | 14.58 % | 1.28 % | 25.66 kcps |
| 147 | 2.23 MBq | $7.68 \pm 1.11\,°\text{C}$ | 3 | 52ph | NE | 12.39 % | 536 ps | 14.67 kcps | 0.09 % | 5.95 % | 0.62 % | 12.95 kcps |
| 148 | 2.23 MBq | $7.68 \pm 1.11\,°\text{C}$ | 3 | 52ph | WE | 12.42 % | 574 ps | 24.53 kcps | 0.12 % | 13.48 % | 0.95 % | 18.33 kcps |
| 149 | 2.15 MBq | $7.66 \pm 1.10\,°\text{C}$ | 2 | 17ph | NE | 12.38 % | 435 ps | 20.51 kcps | 0.07 % | 6.06 % | 0.90 % | 18.07 kcps |
| 150 | 2.15 MBq | $7.66 \pm 1.10\,°\text{C}$ | 2 | 17ph | WE | 12.40 % | 477 ps | 38.92 kcps | 0.10 % | 16.04 % | 1.52 % | 27.39 kcps |
| 151 | 2.08 MBq | $7.67 \pm 1.10\,°\text{C}$ | 2 | 28ph | NE | 12.37 % | 434 ps | 19.30 kcps | 0.07 % | 6.18 % | 0.87 % | 16.96 kcps |
| 152 | 2.08 MBq | $7.67 \pm 1.10\,°\text{C}$ | 2 | 28ph | WE | 12.39 % | 474 ps | 35.63 kcps | 0.10 % | 15.34 % | 1.45 % | 25.49 kcps |

| meas. | activity | system $T_{\text{op}}$ | trig | val | EW | $\Delta E/E$ | CRT | prompts | randoms | scatter | sens | NECR |
|---|---|---|---|---|---|---|---|---|---|---|---|---|
| 153 | 1.85 MBq | 9.12 ± 1.46 °C | 1 | 17ph | NE | 12.53 % | 266 ps | 8.20 kcps | 0.08 % | 5.78 % | 0.42 % | 7.27 kcps |
| 154 | 1.85 MBq | 9.12 ± 1.46 °C | 1 | 17ph | WE | 12.56 % | 288 ps | 15.84 kcps | 0.13 % | 16.79 % | 0.71 % | 10.94 kcps |
| 155 | 1.59 MBq | 9.12 ± 1.48 °C | 1 | 28ph | NE | 12.49 % | 266 ps | 7.55 kcps | 0.07 % | 5.79 % | 0.45 % | 6.69 kcps |
| 156 | 1.59 MBq | 9.12 ± 1.48 °C | 1 | 28ph | WE | 12.51 % | 285 ps | 14.02 kcps | 0.10 % | 15.92 % | 0.74 % | 9.89 kcps |
| 157 | 1.50 MBq | 7.83 ± 1.19 °C | 3 | 28ph | NE | 12.38 % | 539 ps | 13.95 kcps | 0.06 % | 6.05 % | 0.87 % | 12.30 kcps |
| 158 | 1.50 MBq | 7.83 ± 1.19 °C | 3 | 28ph | WE | 12.41 % | 589 ps | 25.72 kcps | 0.10 % | 15.40 % | 1.45 % | 18.38 kcps |
| 159 | 1.41 MBq | 7.70 ± 1.11 °C | 3 | 37ph | NE | 12.36 % | 537 ps | 12.03 kcps | 0.06 % | 5.93 % | 0.80 % | 10.63 kcps |
| 160 | 1.41 MBq | 7.70 ± 1.11 °C | 3 | 37ph | WE | 12.41 % | 582 ps | 21.31 kcps | 0.09 % | 14.72 % | 1.28 % | 15.47 kcps |
| 161 | 1.36 MBq | 7.65 ± 1.10 °C | 3 | 52ph | NE | 12.37 % | 537 ps | 8.91 kcps | 0.06 % | 5.84 % | 0.62 % | 7.89 kcps |
| 162 | 1.36 MBq | 7.65 ± 1.10 °C | 3 | 52ph | WE | 12.38 % | 577 ps | 14.92 kcps | 0.09 % | 13.77 % | 0.95 % | 11.07 kcps |
| 163 | 1.03 MBq | 7.67 ± 1.10 °C | 2 | 17ph | NE | 12.37 % | 434 ps | 9.84 kcps | 0.04 % | 6.04 % | 0.90 % | 8.68 kcps |
| 164 | 1.03 MBq | 7.67 ± 1.10 °C | 2 | 17ph | WE | 12.39 % | 479 ps | 18.73 kcps | 0.07 % | 16.12 % | 1.52 % | 13.16 kcps |
| 165 | 0.99 MBq | 7.66 ± 1.10 °C | 3 | 28ph | NE | 12.36 % | 434 ps | 9.28 kcps | 0.05 % | 6.15 % | 0.88 % | 8.16 kcps |
| 166 | 0.99 MBq | 7.66 ± 1.10 °C | 3 | 28ph | WE | 12.40 % | 475 ps | 17.31 kcps | 0.08 % | 15.91 % | 1.47 % | 12.23 kcps |
| 167 | 0.92 MBq | 8.86 ± 1.37 °C | 1 | 17ph | NE | 12.52 % | 266 ps | 4.18 kcps | 0.05 % | 6.02 % | 0.43 % | 3.69 kcps |
| 168 | 0.92 MBq | 8.86 ± 1.37 °C | 1 | 17ph | WE | 12.53 % | 287 ps | 8.09 kcps | 0.09 % | 16.32 % | 0.73 % | 5.65 kcps |
| 169 | 0.88 MBq | 9.00 ± 1.44 °C | 1 | 28ph | NE | 12.53 % | 267 ps | 4.23 kcps | 0.04 % | 6.00 % | 0.45 % | 3.73 kcps |
| 170 | 0.88 MBq | 9.00 ± 1.44 °C | 1 | 28ph | WE | 12.54 % | 285 ps | 7.86 kcps | 0.08 % | 15.86 % | 0.75 % | 5.56 kcps |
| 171 | 0.80 MBq | 7.79 ± 1.15 °C | 3 | 28ph | NE | 12.36 % | 539 ps | 7.46 kcps | 0.04 % | 6.10 % | 0.88 % | 6.57 kcps |
| 172 | 0.80 MBq | 7.79 ± 1.15 °C | 3 | 28ph | WE | 12.40 % | 592 ps | 13.77 kcps | 0.07 % | 15.72 % | 1.45 % | 9.77 kcps |
| 173 | 0.75 MBq | 7.68 ± 1.11 °C | 3 | 37ph | NE | 12.34 % | 537 ps | 6.44 kcps | 0.04 % | 5.98 % | 0.80 % | 5.69 kcps |
| 174 | 0.75 MBq | 7.68 ± 1.11 °C | 3 | 37ph | WE | 12.43 % | 586 ps | 11.45 kcps | 0.07 % | 16.19 % | 1.27 % | 8.03 kcps |
| 175 | 0.72 MBq | 7.64 ± 1.09 °C | 3 | 52ph | NE | 12.37 % | 536 ps | 4.74 kcps | 0.04 % | 5.90 % | 0.62 % | 4.19 kcps |
| 176 | 0.72 MBq | 7.64 ± 1.09 °C | 3 | 52ph | WE | 12.42 % | 579 ps | 7.95 kcps | 0.07 % | 15.62 % | 0.93 % | 5.65 kcps |
| 177 | 0.68 MBq | 7.66 ± 1.10 °C | 2 | 17ph | NE | 12.35 % | 433 ps | 6.50 kcps | 0.03 % | 6.66 % | 0.89 % | 5.66 kcps |
| 178 | 0.68 MBq | 7.66 ± 1.10 °C | 2 | 17ph | WE | 12.39 % | 480 ps | 12.38 kcps | 0.06 % | 16.48 % | 1.53 % | 8.63 kcps |
| 179 | 0.64 MBq | 7.66 ± 1.09 °C | 2 | 28ph | NE | 12.34 % | 434 ps | 6.02 kcps | 0.03 % | 6.21 % | 0.87 % | 5.29 kcps |

Table S3 – continued from previous page

| meas. | activity | system $T_\mathrm{op}$ | trig | val | EW | $\Delta E/E$ | CRT | prompts | randoms | scatter | sens | NECR |
|---|---|---|---|---|---|---|---|---|---|---|---|---|
| 180 | 0.64 MBq | 7.66 ± 1.09 °C | 2 | 28ph | WE | 12.42 % | 479 ps | 11.17 kcps | 0.06 % | 15.69 % | 1.46 % | 7.93 kcps |
| 181 | 0.61 MBq | 7.58 ± 1.08 °C | 3 | 28ph | NE | 12.36 % | 540 ps | 5.70 kcps | 0.04 % | 6.24 % | 0.87 % | 5.01 kcps |
| 182 | 0.61 MBq | 7.58 ± 1.08 °C | 3 | 28ph | WE | 12.41 % | 597 ps | 10.60 kcps | 0.07 % | 16.18 % | 1.46 % | 7.44 kcps |
| 183 | 0.58 MBq | 7.63 ± 1.08 °C | 3 | 37ph | NE | 12.33 % | 536 ps | 4.95 kcps | 0.04 % | 5.94 % | 0.80 % | 4.38 kcps |
| 184 | 0.58 MBq | 7.63 ± 1.08 °C | 3 | 37ph | WE | 12.39 % | 588 ps | 8.85 kcps | 0.07 % | 15.17 % | 1.29 % | 6.36 kcps |
| 185 | 0.55 MBq | 7.63 ± 1.09 °C | 3 | 52ph | NE | 12.37 % | 535 ps | 3.61 kcps | 0.04 % | 5.77 % | 0.62 % | 3.21 kcps |
| 186 | 0.55 MBq | 7.63 ± 1.09 °C | 3 | 52ph | WE | 12.43 % | 578 ps | 6.08 kcps | 0.07 % | 14.01 % | 0.95 % | 4.49 kcps |